\documentclass[prb,aps,amssymb,showpacs,twocolumn]{revtex4-1}
\usepackage{bm,color,amsmath,amssymb,mathrsfs,latexsym,graphicx,psfrag}

\newcommand{\beq}{\begin{equation}}
\newcommand{\eneq}{\end{equation}}

\input{epsf}

\begin{document}

\tolerance 10000

\newcommand{\vk}{{\bf k}}
\def\ns{^{\vphantom{*}}}
\def\ket#1{{|  #1 \rangle}}
\def\bra#1{{\langle #1|}}
\def\braket#1#2{{\langle #1  |   #2 \rangle}}
\def\expect#1#2#3{{\langle #1 |   #2  |  #3 \rangle}}
\def\cH{{\cal H}}
\def\half{\frac{1}{2}}
\def\sut{\textsf{SU}(2)}
\def\suto{\textsf{SU}(2)\ns_1}
\def\kF{\ket{\,{\rm F}\,}}


\title{Decomposition of fractional quantum Hall states: New symmetries
  and approximations}

\author{Ronny Thomale$^{1}$, Benoit Estienne$^{2}$, Nicolas Regnault$^{3}$, and  B. Andrei Bernevig$^{1}$}

\affiliation{$^1$ Department of Physics, Princeton University,
  Princeton, NJ 08544, USA}
\affiliation{$^2$ Institute for Theoretical Physics, Universiteit van Amsterdam Valckenierstraat 65, 1018 XE Amsterdam, The Netherlands}
\affiliation{$^3$ LPA, Departement de Physique, ENS, CNRS, 24 rue Lhomond,
  75005 Paris, France}

\begin{abstract}
We provide a detailed description of a new symmetry structure of the
monomial (Slater) expansion coefficients of bosonic (fermionic)
fractional quantum Hall states first obtained in
Ref.~\onlinecite{bernevig-09prl206801}, which we now extend to
spin-singlet states. We show that the Haldane-Rezayi spin-singlet
state can be obtained without exact diagonalization through a
differential equation method that we conjecture to be generic to
other FQH model states. The symmetry rules in
Ref.~\onlinecite{bernevig-09prl206801} as well as the ones we obtain
for the spin singlet states allow us to build approximations of FQH
states that exhibit \emph{increasing} overlap with the exact state
(as a function of system size). We show that these overlaps reach
unity in the thermodynamic limit even though our approximation omits
more than half of the Hilbert space. We show that the product rule is
valid for any FQH state which can be written as an expectation value
of parafermionic operators.
\end{abstract}

\date{\today}


\maketitle

\section{Introduction}

Our understanding of the physics of the fractional quantum Hall effect
(FQHE) has benefited greatly from the existence of model wave
functions. Laughlin's trial wave function for the $\nu=1/3$ filled FQH
state provided the first paradigm to understand the emergent behavior
of interacting electrons in a strong magnetic
field~\cite{laughlin83prl1395}. The current understanding of trial
wave functions predominantly uses the conformal field theory (CFT)
connection first proposed in Ref.~\onlinecite{Moore-91npb362}. For
every existing CFT, one can build a FQH trial wave function by taking
the expectation value of branch-cut free primary fields in the CFT.
The Read-Rezayi (RR) states are a product of this line of
reasoning~\cite{Read-99prb8084}. Both spin-polarized as well as
spin-singlet states can be obtained this way, most prominent examples
of which are the Haldane-Rezayi (HR), the Non-Abelian spin singlet
(NASS), and the Halperin
states~\cite{haldane-88prl956,halperin83hpa75,ardonne-99prl5096}.  FQH
trial wave functions are essential to understanding the physically
important concepts of fractional Abelian (in the Laughlin and
composite fermion
states~\cite{arovas-84prl722,halperin84prl1583,jain89prl199}) and
non-Abelian statistics (in the Moore-Read (MR)~\cite{Moore-91npb362}
and RR states~\cite{Read-99prb8084}).

The central drawback of the CFT-motivated trial wave function approach
is the lack of both an explicit decomposition of a trial state in a
second quantized many-body basis and of a first quantized closed form
expression for the state. As a consequence, Monte Carlo methods, while
useful for Laughlin states~\cite{laughlin83prl1395}, cannot be applied
for most non-Abelian states.
Any quantitative analysis of these trial states has hence so far
relied on exact diagonalization (ED)
methods~\cite{haldane83prl605,haldane-85prl237}.
In these methods, one starts with a trial Hamiltonian and generates
the (lowest) Landau level (LLL) Hilbert space. The computational
effort of diagonalization depends algebraically on the Hilbert space
dimension, which grows factorially with system size. This sets the
size limit that is reachable from ED. It is hence essential
to use all available symmetries contained in the trial state and in
the associated trial Hamiltonian to find the smallest subblock
structure of the Hamiltonian matrix in terms of the non-interacting
basis. One symmetry is the reflection of angular momentum $L_z
\rightarrow - L_z$ which, for a sphere geometry, is equivalent to the
indistinguishability of the north and south pole. Other symmetries
such as total $L^2$ multiplet structure exist in some cases. 
However, they are rather obvious in general and do not gain us deep
insight in the structure of the FQH states.  
For the Laughlin $1/3$ state, previous
attempts~\cite{dunne93ijmp4783,difrancesco-94ijmp4257} to calculate
the weights of the free many-body wave functions in the full
interacting state failed. These
works~\cite{dunne93ijmp4783,difrancesco-94ijmp4257} obtain only
$O(1/N!)$ of the $O(N!)$ coefficients, and hence represent a set of
measure zero of the Laughlin state expansion.  In a recent paper, two
of us found that a large series of FQH trial states obey a new type of symmetry for
their free basis expansion coefficients~\cite{bernevig-09prl206801}.
The symmetry relates a subset of the coefficients of
the expansion in free many-body states of a given FQH state to products
of state coefficients from smaller system size.  This was developed
for bosonic and spin-polarized fermionic
states. In particular, it was observed in Ref.~\onlinecite{bernevig-09prl206801}
that the overlap of the exact FQH state with the state approximated by
the "product rule" symmetry increases with system size asymptotically
towards unity.

In this paper, we give a detailed account of a general differential
equation method used in Ref.~\onlinecite{bernevig-09prl206801} to access
the monomial (Slater) decomposition of bosonic (fermionic) FQH states.
We provide a detailed description of the very condensed derivation in Ref.~\onlinecite{bernevig-09prl206801} of the expansion coefficients for bosonic and polarized
fermionic states. From there, we explain how the trial state can be
numerically generated at a level intended for the novice reader. 
Next, we present an extended proof of the product rule symmetry for
FQH trial states (and for all Jack polynomials) previously summarized
in Ref.~\onlinecite{bernevig-09prl206801}.  We then extend the product
rule symmetry, which allows to generate even
more expansion coefficients than previously allowed. We also
generalize the whole approach to spinful trial states, and illustrate
it in detail for the HR state.  We first derive an annihilation
operator for the HR state from which we develop a
recurrence relation for the expansion coefficients.  We
investigate the product rule symmetry analogue for spinful states and
extract the entanglement spectrum of the HR state.  For a spinful
trial state, we find that the particle number $N$, angular momentum $L$, and
spin multiplet $S$ are the quantum numbers of the reduced density
matrix subblocks.

The article is organized as follows. In Section~\ref{sec:boson}, we
discuss the recurrence relation of Jack polynomials that leads to the
monomial expansion coefficients for bosonic FQH trial states.  We
elaborate on numerical subtleties for certain negative Jack parameters
$\alpha$. In these cases, denominator divergences appear in the
recurrence formula: they are accompanied by an (at least) similarly
vanishing numerator. In Section~\ref{sec:fermion}, the Slater
expansion coefficients of spin-polarized fermionic FQH states are
derived from a fermionic version of the Laplace-Beltrami operator. This
is the expanded version of previous calculations presented
in Ref.~\onlinecite{bernevig-09prl206801}. The approach is used to develop a
recurrence formula for fermionic FQH trial states.  
In Section~\ref{sec:product}, we provide a largely expanded proof of
the product rule symmetry.  For non-Abelian bosonic states, we extend
the product rule to treat general cases of cutting through a multiply
occupied root partition orbital. In Section~\ref{sec:hr}, we
generalize the entire approach to the spinful HR state. We derive the
recurrence formula, show the product rule property, and compute the
entanglement spectrum of this spinful trial state. In
Section~\ref{sec:cft}, we take a general viewpoint on the product rule
symmetry from conformal field theory. We show that the product rule
manifests itself as a generic property of all FQH states which can be
written as an expectation value of parafermionic operators hence
including a large set of both spin polarized and spin unpolarized FQH states.
Finally, we conjecture in Section~\ref{sec:con} that the product rule
symmetry is a structural property of the majority of FQH trial states
including fermionic or bosonic states and spin-polarized or
spin-unpolarized states, and potentially serves as an important
ingredient to density matrix renormalization group approaches for FQH
systems.

\section{Bosonic states}
\label{sec:boson}

FQH states are analytic functions of the positions of electrons in a
magnetic field. The single-particle orbitals in the Landau Level are
given by $\phi_m(z)=(2\pi m!  2^m)^{-1/2} z^m \exp (-\vert
z\vert^2/4)$ with angular momentum $L_z=m\hbar$, although from now we
will neglect the trivial Gaussian multiplication factors. A
non-interacting N-particle basis state can be indexed by a partition
$\lambda$ - an ordered list of the $L_z$ angular momentum of the occupied orbitals.
 The corresponding occupation number configuration is
$n(\lambda)=\{n_m(\lambda),m=0,1,2,\dots\}$~\cite{haldane06baps633,bernevig-08prl246802},
where $m$ labels the individual single-particle orbitals and
$n_m(\lambda)$ is the multiplicity of orbital $m$ in $\lambda$.  We
consider FQH states decomposed in this many-body basis, either of
bosons (permanents) or fermions (slaters) with expansion
coefficients $c_{\lambda}$. The central task of this paper is to
develop methods to compute these expansion coefficients.

\begin{figure}[t]
  \begin{minipage}[l]{0.65\linewidth}
    \includegraphics[width=\linewidth]{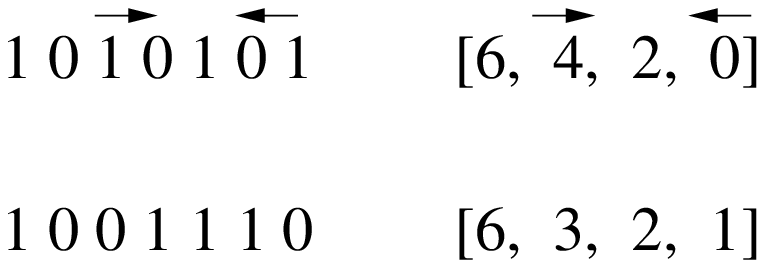}
  \end{minipage}
  \caption{Pictorial example of the squeezing operation in occupation
    language (left) and partition language (right). The squeezing
    operation takes the first row into the second.}
\label{squeez}
\vspace{-0pt}
\end{figure}

We now define a two-body operation on the many-body basis that is
important for the purpose of the paper: for a pair of particles in the
orbitals $m_1$ and $m_2$, with $m_1<m_2-1$, the elementary {\it
  squeezing} operation consists of the two particles shifted to
different momentum orbitals as $n_{m_{1,2}}\rightarrow n_{m_{1,2}}-1$,
$n_{m_{1,2}\pm 1}\rightarrow n_{m_{1,2}\pm 1}+ 1$.  This means that
both particles in the $m_1, m_2$ orbitals are shifted "inwards" the
partition (as shown in Fig.~\ref{squeez}).
\noindent
The squeezing defines a partial ordering relation between two
partitions $\lambda > \mu$ when $\mu$ is generated by squeezing
operations acting on
$\lambda$~\cite{stanley89am76,sutherland71pra2019}. This ordering
yields a tree hierarchy a complete example of which is shown in
Appendix~\ref{app:mono}.  By contrast, when $\lambda$ and $\mu$ do not
relate by squeezing, no ordering relation is set between these
partitions.

The trial FQH states we consider are all squeezed polynomials. They
possess a unique partition, called the root partition, dominating all
other partitions. This means that all partitions with possible (but
not guaranteed) non-zero weight are generated by subsequent squeezing
operations acting on the root partition. In many cases, this already
allows us to omit a significant (more than half) part of the Hilbert space (see
Table~\ref{tab:dim}).

\begin{table}[t]
\begin{center}
\begin{tabular}{c|c|c}
nbr particles & full dim. & squeezed dim. \\
\hline
4 & 18 & 16\\
5 & 73 & 59\\
6 & 338 & 247\\
7 & 1656 & 1111\\
8 & 8512 & 5302\\
9 & 45207 & 26376\\
10 & 246448 & 135670\\
11 & 1371535 & 716542\\
12 & 7764392 & 3868142\\
13 & 44585180 & 21265884\\
14 & 259140928 & 118741369\\
15 & 1521967986 & 671906876\\
16 & 9020077206 & 3846342253\\
17 & 53885028921 & 22243294360
\end{tabular}
\end{center}
\caption{Size of the monomial basis for the bosonic Laughlin state
  $\nu=1/2$ up to $N=17$ particles. The second column is the complete
  size. The third column is the number of partitions allowed by the
  squeezing operation from the root partition $1010101 \dots 0101$. }
\label{tab:dim}
\end{table}

In this Section, we focus on the bosonic FQH states. The
non-interacting basis is given by monomials
$\mathcal{M}_\lambda(z_1,\dots, z_N) = \text{Per}
(z_i^{\lambda_j})/\prod_m n_m(\lambda)!$, where $\lambda_j$ is
 the momentum index of the $j$th particle in the partition
$\lambda$ and $\text{Per}$ is the permanent. It was
shown~\cite{bernevig-08prl246802} that the $N$-particle bosonic RR $k$
series of states (which includes the Laughlin and MR state) are a
special class of symmetric polynomials.  Specifically, this class is
called the $r=2$ single \emph{Jack} polynomials $J_\lambda^\alpha
(z_1,\dots,z_N)$ of parameter $\alpha =-\frac{k+1}{r-1}$ and root
partition $\lambda=[k0^{r-1}k\dots k0^{r-1}k]$. The Jack wave
functions can be related to $\text{WA}_{k-1}$ conformal field theories
and can be classified in terms of symmetric polynomial
categories~\cite{estienne-jpa445209,bernevig-09jpa245206,mathieu-cm10,jolicoeur-cm10,mathieu09jmp095210,estienne-10npb539,santachiara-cm10}.
Moreover, the quasi-particle excitations of the trial state systems
can also be written as coherent state superpositions of
Jacks~\cite{bernevig-09prl066802,estiennearx1005}.  This provides a complementary view
to that of other approaches for FQH quasi-particle excitation
states~\cite{hansson-09prl166805,wen-08prb155109,barkeshli-09prb195132,ardonne-08jsm04016,ardonne09prl180401,lu-10prb115124,seidel08prl196802,seidel-cm10,bergholtz-08prb155308,seidel-08prl036804,regnault-08prl066803,wojscm08,read09prb045308,read09prb245304,hanssonka-cm09,wikberg-09jsm07038,wojs09prb041104,papic-09prb201303,toke-09prb205301,estiennearx1005,degierarx1007}.

Jacks are eigenstates of the Laplace Beltrami (LB)
operator~\cite{stanley89am76}:
\begin{equation}
\mathcal{H}_{\text{LB}}=\sum_i \left( z_i \frac{\partial}{\partial z_i} \right)^2 + \frac{1}{\alpha} \sum_{i<j} \frac{z_i+z_j}{z_i-z_j} \left( z_i \frac{\partial}{\partial z_i} -z_j \frac{\partial}{\partial z_j}\right).
\end{equation} 
Until recently, the spectrum of the LB operator had been studied
in detail for only positive $\alpha$ in the context of the
Calogero-Sutherland model~\cite{sutherland71pra2019}. Recent progress
has shown that the LB operator has well-defined polynomial solutions
for certain negative $\alpha$~\cite{feigin-02imrn1223}. In particular,
as stated above, some of them are found to correspond to bosonic FQH
trial states for the ground state, quasi-electron and quasi-hole
excitations~\cite{bernevig-08prb184502}.

We expand the Jacks into the monomial basis :
\begin{equation}
J_{\lambda}^{\alpha} = \sum_{\kappa \le \lambda} c_{\lambda \kappa}(\alpha) \mathcal{M}_\kappa,
\end{equation}
where $\kappa$ runs over all monomial partitions squeezed from the root partition $\lambda$.
There is a known recurrence
relation for the expansion coefficients $ c_{\lambda
  \kappa}(\alpha)$~\cite{dumitriu-07jsc,ha95npb604}:
\begin{equation}
c_{\lambda \kappa}(\alpha) = \frac{2/\alpha}{\rho_{\lambda}(\alpha)-\rho_{\kappa}(\alpha)} \sum_{\kappa<\mu\le \lambda}\hspace{-5pt} \left( (l_i+t)-(l_j-t) \right) c_{\mu \kappa}(\alpha), \label{recbos}
\end{equation}
where $\kappa=[l_1,\dots,l_i,\dots,l_j,\dots]$ and
$\mu=[l_1,\dots,l_i+t,\dots,l_j-t,\dots]$ denote partitions. We
arrange the momentum orbitals denoted above in decreasing order from
left to right, i.e. $l_1 \ge l_2 \ge l_i \ge l_j \dots$ in $\kappa$,
and a possible rearrangement occurs in $\mu$ depending on $t$.  All
partitions $\mu$ are understood to be reordered in this way. The sum
in~\eqref{recbos} extends over all partitions $\mu$ strictly
dominating $\kappa$ but being dominated (squeezed from) or equal to
$\lambda$ that can be generated by unsqueezing (i.e. the inverse
operation to squeezing). The $\rho$'s are defined as:
\begin{equation}
\rho_{\lambda}(\alpha)=\sum_i \lambda_i \left(\lambda_i -1 - \frac{2}{\alpha}(i-1)\right).\label{btof}
\end{equation}

We now explain an easily implementable computer algorithm that allows
one to obtain a large number of bosonic FQH states with high
precision. From~\eqref{btof} we can uniquely index every partition by
$\sum_i 2^{\lambda_i+N-i}$. For any numerical implementation, we order
the basis according to this index, so that the look-up of a partition
in the basis list becomes logarithmic in effort. By recurrence, we can
always compute the coefficient of a partition from those coefficients
of the partitions dominating it. The number of dominating partitions
is small.  Averaged over all partitions, it approximately scales with
the number of fluxes ($\sim$ number of orbitals) times the square of the number of
particles i.e. $N^2 N_\phi$.
 Thus, to compute all expansion coefficients, the
procedure gives linear effort in the monomial basis dimension. The
algorithm to generate the Jack state is sketched in
Table~\ref{tab:howtorec}.
\begin{widetext}
\begin{table*}[t]
\begin{center}
\begin{tabular}{cc}
  \hline\hline
  &Algorithmic steps to generate the Jack state \\
  \hline
  (i) &  Generate squeezed monomial basis $\mathcal{M}_\kappa$; $\mathcal{M}_1=\mathcal{M}_\lambda$ is the root partition\\
  &  Order basis states by the integers $\sum_i 2^{\lambda_i+N-i}$\\
  (ii)&  Compute all $\rho_\kappa (\alpha)$ (Eq.~\ref{btof})\\
  (iii)&  Loop over all $\kappa$: Loop over all pairs of elements $l_i < l_j \in \kappa$;\\ 
  & For each pair unsqueeze to upper dominant partitions $\mu>\kappa$ and read off $c_{\lambda \mu}$ \\
  (iv)&  Compute contribution to $c_{\lambda \kappa}$ by Eq.~\ref{recbos};\\
  &  if $\rho_{\kappa}(\alpha)=\rho_{\lambda}(\alpha)$ compute the limit prescription $\text{lim}_{\epsilon \rightarrow 0} [\alpha \rightarrow \alpha-\epsilon]$\\

\hline\hline 
\end{tabular}
\end{center}
\caption{Sketched howto to use Eq.~\ref{recbos} to generate bosonic FQH states in monomial basis and Jacks of arbitrary parameter $\alpha$.}
\label{tab:howtorec}
\vspace{-5pt}
\end{table*}
\end{widetext}
In contrast to positive $\alpha$ for which this formula was originally
derived, there are minor caveats for certain negative $\alpha$.
Situations occur in which the denominator in~\eqref{recbos} vanishes
for certain partitions. In these cases, one can find that two different partitions
$\mu_1$ and $\mu_2$ obey $\rho_{\mu_1}(\alpha)=\rho_{\mu_2}(\alpha)$,
a situation that can be proved to never arise for positive $\alpha$.
An elementary example would be the $6$-particle partitions
$\mu_1(-4/3)=[5,5,4,1,1,0]$ and $\mu_2(-4/3)=[3,3,3,3,2,2]$. However,
this denominator divergence is always regularized by a vanishing
numerator. Under a limiting prescription $\text{lim}_{\epsilon
  \rightarrow 0} [\alpha \rightarrow \alpha-\epsilon]$, the quotient
either gives $0$ or a rational value.  Numerically, we let $\alpha$
slightly deviate from its exact value, vary it, and find a plateau
value, which is then identified as the resulting expansion coefficient
for the partition. This type of $\rho$ degeneracy does happen neither for
the $r=2$ RR series nor for the Gaffnian
state~\cite{read09prb045308,simon-07prb075317,simon-07prb075318}. It
only occurs for Jacks that were shown not to be uniquely defined by a single clustering
condition~\cite{bernevig-08prb184502}.



\section{Polarized fermionic states}
\label{sec:fermion}
Similar to the bosonic case in Sec.~\ref{sec:boson}, we start with
single particle orbitals of the Landau level defined before, i.e.
$\phi_m(z)=(2\pi m!  2^m)^{-1/2} z^m \exp (-\vert z\vert^2/4)$ with
angular momentum $L_z=m\hbar$ and $m$ the labeling index for all
single particle orbitals. However, for the many-body state, we now
assume that the particles described by the first quantized wave
functions obey fermionic statistics. As a consequence, the
non-interacting free fermion basis is given by Slater determinant
states: $\text{sl}_\lambda=\mathcal{A}_z (z_1^{\lambda_1}
z_2^{\lambda_2} \dots z_N^{\lambda_N}) = \text{Det}(z_i^{\lambda_j}).$
$\text{sl}_\lambda$ is the unnormalized orthogonal Slater determinant,
where $\mathcal{A}$ denotes the antisymmetrization over all $z$
coordinates. Different normalizations can be applied to put the
polynomial wave function on different manifolds such as the plane or
the sphere.
As in the bosonic case, we describe the free many body states by
partitions (or occupation numbers).  We again assume the partition
$\lambda=[\lambda_1,\dots, \lambda_N]$ to be ordered by decreasing
order in angular momentum $\lambda_i$ of the $i$th particle. As
before, the squeezing operation shifts two particles inwards (towards
each other) in the partition. For fermions, multiple occupancy is
forbidden due to the Pauli principle.

In first quantized notation, bosonic and fermionic trial states can be
transformed into each other by multiplication with a Vandermonde
determinant. In terms of single particle coordinates, this polynomial
is the Jastrow factor, which is the antisymmetric homogeneous
polynomial of degree $1$.  Starting from a Jack polynomial
$J_{\lambda}^{\alpha}$, the transformation reads
$J_{\lambda}^{\alpha}\rightarrow S_{\lambda}^{\alpha} :=
J_{\lambda}^{\alpha}\prod_{i<j}(z_i-z_j)$. The $S_{\lambda}^{\alpha}$
polynomials are the exact fermionic analogue of the bosonic (Jack)
trial state $J_{\lambda}^\alpha$. For example, the $\nu=1/2$ bosonic
Laughlin state (Jack of $(k,r)=(1,2)$) becomes the $\nu=1/3$ fermionic
Laughlin state. The filling always changes from bosonic filling
$\nu=p/q$ to fermionic filling $\nu=p/(p+q)$.  However, in second
quantized notation, multiplication by the Vandermonde determinant does
not transform a single monomial to a single Slater.  To obtain a
one-to-one correspondence between a bosonic basis and fermionic
Slaters, one would first have to transform the monomials to Schur
functions~\cite{schur06sawb}.  However, this involves knowledge of all
the Kostka numbers, a long-standing unsolved mathematical problem with
no known efficient algorithm~\cite{kostka82jam}. There are two ways to
remedy this problem. First, we can use the
knowledge that the transformation from monomials to Schur functions is
exactly given by the $J^{\alpha=1}$ Jack polynomial coefficients,
which we can compute from Eq.~\ref{recbos}.

However, we try
to tackle the fermionic trial states from a different angle. We define
the decomposition of the $S_{\lambda}^{\alpha}$ polynomials into
Slaters:
\begin{equation}
S_{\lambda}^{\alpha}(z_1,\dots,z_N)=J_{\lambda_{\text{B}}}^\alpha \prod_{i<j}^N (z_i-z_j)=\sum_{\mu \le \lambda} b_{\lambda \mu} \text{sl}_\mu. \label{salpha}
\end{equation}
To avoid confusion, $\lambda_{\text{B}}$ denotes the bosonic root
partition and $\lambda$ the associated fermionic root partition.  All
partitions $\mu$ are squeezed from the fermionic partition $\lambda$ that is
related to the bosonic partition by $\lambda_i=\lambda_i^B + (N-i)$.
We now use that the Jack part of $S_{\lambda}^{\alpha}$ is an
eigenstate of the LB operator, i.e.
$\mathcal{H}_{\text{LB}}J_{\lambda_{\text{B}}}^\alpha=E_{\lambda_{\text{B}}}(\alpha)J_{\lambda_{\text{B}}}^\alpha$.
We then relate the derivatives acting on
$J_{\lambda_{\text{B}}}^\alpha$ to derivatives on
$S_{\lambda}^{\alpha}$ (details are given in
Appendix~\ref{app:flb}):
\begin{widetext}
\begin{eqnarray}
E_{\lambda_{\text{B}}}(\alpha)S_{\lambda}^{\alpha}&=&\prod_{k<l} (z_k-z_l) \left[\sum_i \left(z_i \frac{\partial}{\partial z_i}\right)^2 +\frac{1}{\alpha} \sum_{i<j} \frac{z_i+z_j}{z_i-z_j} \left( z_i \frac{\partial}{\partial z_i} - z_j \frac{\partial}{\partial z_j}\right) \right] J_{\lambda_{\text{B}}}^\alpha \nonumber \\
&=&\left[ \sum_i \left(z_i \frac{\partial}{\partial z_i}\right)^2-2\sum_{\substack{i,m \\ m\ne i}} \frac{z_i}{z_i-z_m}z_i\frac{\partial}{\partial z_i}+\frac{1}{\alpha} \sum_{i<j} \frac{z_i+z_j}{z_i-z_j} \left( z_i \frac{\partial}{\partial z_i} - z_j \frac{\partial}{\partial z_j}\right)\right]S_{\lambda}^{\alpha} \nonumber \\
&& +\left[\sum_{\substack{i,m\\i\ne m}} \frac{z_i (z_i+z_m)}{(z_i-z_m)^2} + \sum_{\substack{i,m,n\\i\ne m \ne n}} \frac{z_i^2}{(z_i-z_m)(z_i-z_n)}+ \frac{1}{\alpha} \sum_{i<j} \frac{z_i+z_j}{z_i-z_j}\left(\sum_{m\ne i} \frac{z_i}{z_i-z_m}- \sum_{m\ne j} \frac{z_j}{z_j-z_m} \right)\right]S_{\lambda}^{\alpha}.\nonumber \\
\end{eqnarray}
Simplifying several polynomial sums that yield constants as shown in
Appendix~\ref{app:form}, we can define a fermionic Laplace Beltrami
operator that diagonalizes $S_\lambda^\alpha$, i.e.
$\mathcal{H}^{\text{F}}_{\text{LB}}(\alpha)S_{\lambda}^{\alpha}(z_1,\dots,z_N)=E_\lambda
(\alpha)S_{\lambda}^{\alpha}(z_1,\dots,z_N)$, with
\begin{equation}
E_{\lambda}(\alpha) = \sum_i \lambda_i \left( \lambda_i -2 \left( \frac{1}{\alpha}-1 \right) i \right) + \left( \frac{1}{\alpha} -1\right) \left( \left( N+1 \right) \sum_i \lambda_i - N \left( N-1 \right) \right),
\end{equation} 
\begin{equation}
\mathcal{H}_{\text{LB}}^{\text{F}}(\alpha)= H_K + \frac{1}{2}\left(\frac{1}{\alpha}-1 \right) H_I = \sum_i \left(z_i \frac{\partial}{\partial z_i} \right)^2 + \frac{1}{2}\left (\frac{1}{\alpha}-1 \right)\left[ \sum_{i \ne j} \frac{z_i+z_j}{z_i-z_j} \left(z_i \frac{\partial}{\partial z_i} - z_j \frac{\partial}{\partial z_j}\right) -2 \frac{z_i^2+z_j^2}{(z_i-z_j)^2}\right].
\end{equation}
\end{widetext}
We now diagonalize the above operator in the basis of Slater
determinants. The action of the kinetic part yields $\sum_i H_K
\text{sl}_\mu = (\sum_i \mu_i^2) \text{sl}_\mu$, where the $\mu_i$
denotes the polynomial power of the $i$th particle in the partition.
The action of the interaction part $H_I$ is non-diagonal in Slater
determinant basis and demands further calculation. First we realize
that, due to its two-body nature, the action of $H_I$ on any Slater
determinant decomposes into the sum of two-particle interaction
terms. It is thus sufficient to look at the action on the two-particle
Slater determinant
$\text{sl}_{\mu=(\mu_1,\mu_2)}=z_1^{\mu_1}z_2^{\mu_2}-z_2^{\mu_1}z_1^{\mu_2}$.
Assume $\mu_1 > \mu_2$, and define $p=\mu_1-\mu_2$:
\begin{widetext}
\begin{eqnarray}
&& \frac{H_I \text{sl}_{(\mu_1,\mu_2)}}{z_1^{\mu_2}z_2^{\mu_2}}= p \frac{z_1+z_2}{z_1-z_2} (z_1^{p} + z_2^{p}) -2 \frac{z_1^2+z_2^2}{(z_1-z_2)^2} (z_1^{p}-z_2^{p})\nonumber\\
&=& \frac{1}{z_1-z_2} \Big( p (z_1^{p+1}+z_2^{p+1}+z_1^p z_2 + z_2^p z_1 ) -2\sum_{s=1}^{p/2} \left(z_1^{p-s} z_2^{s+1} + z_2^{p-2}z_1^{s+1} + z_1^{p-s+2}z_2^{s-1}+z_2^{p-s+2}z_1^{s-1} \right)\Big) \nonumber \\
&=& \frac{1}{z_1-z_2} 2 \sum_{s=1}^{p/2} \Big( z_1^{p-s+2} (z_1^{s-1}-z_2^{s-1}) +z_2^{p-s+2} (z_2^{s-1}-z_1^{s-1}) + z_1^{p-s}z_2 (z_1^s-z_2^s) + z_2^{p-s} z_1 (z_2^s-z_1^s)\Big)\nonumber \\
&=&  2 \sum_{s=1}^{p/2} (z_1^{p-s+2}-z_2^{p-s+2}) \sum_{t=1}^{(s-1)/2} (z_1^{s-t-1}z_2^{t-1} + z_2^{s-t-1}z_1^{t-1})+ 2 \sum_{s=1}^{p/2} (z_1^{p-s}z_2-z_2^{p-s}z_1) \sum_{t=1}^{s/2}(z_1^{s-t}z_2^{t-1}+z_2^{s-t}z_1^{t-1})
\end{eqnarray}
The two terms are already grouped to yield two-particle Slater
determinants. Collecting all prefactors, this gives:
\begin{equation}
H_I \text{sl}_{(\mu_1,\mu_2)}= (\mu_1-\mu_2-2)\text{sl}_{(\mu_1,\mu_2)} +2 \sum_{s=1}^{(\mu_1-\mu_2)/2} (\mu_1-\mu_2-2s) \text{sl}_{(\mu_1-s,\mu_2+s)}.
\label{slater1}
\end{equation}
\end{widetext}
Eq.~\ref{slater1} has a particular form: it only scatters "inwards"
the two-particle basis of Slater determinants, i.e. towards decreasing
relative momentum of the particles, and thus to a squeezed partition.
Let us now look at the total action of $H_{\text{LB}}^{\text{F}}$ on
$S^\alpha_\lambda$ expanded in Slaters.  The above scattering
Hamiltonian and the linear independence of Slater determinants provide
a recurrence relation for the coefficients $b_{\lambda \mu}$
in~\eqref{salpha}. We collect all diagonal terms and invert the sum
over $s$ in Eq.~\ref{slater1} to a sum over all dominating partitions:
\begin{equation}
b_{\lambda \mu} = \frac{2(\frac{1}{\alpha}-1)}{\rho^{\text{F}}_\lambda (\alpha) - \rho^{\text{F}}_\mu (\lambda)} \sum_{\theta; \; \mu < \theta \le \lambda} (\mu_i-\mu_j) b_{\lambda \theta} (-1)^{N_{\text{SW}}},
\label{slater2}
\end{equation}

\begin{figure}[t]
  \begin{minipage}[l]{0.99\linewidth}
    \includegraphics[width=\linewidth]{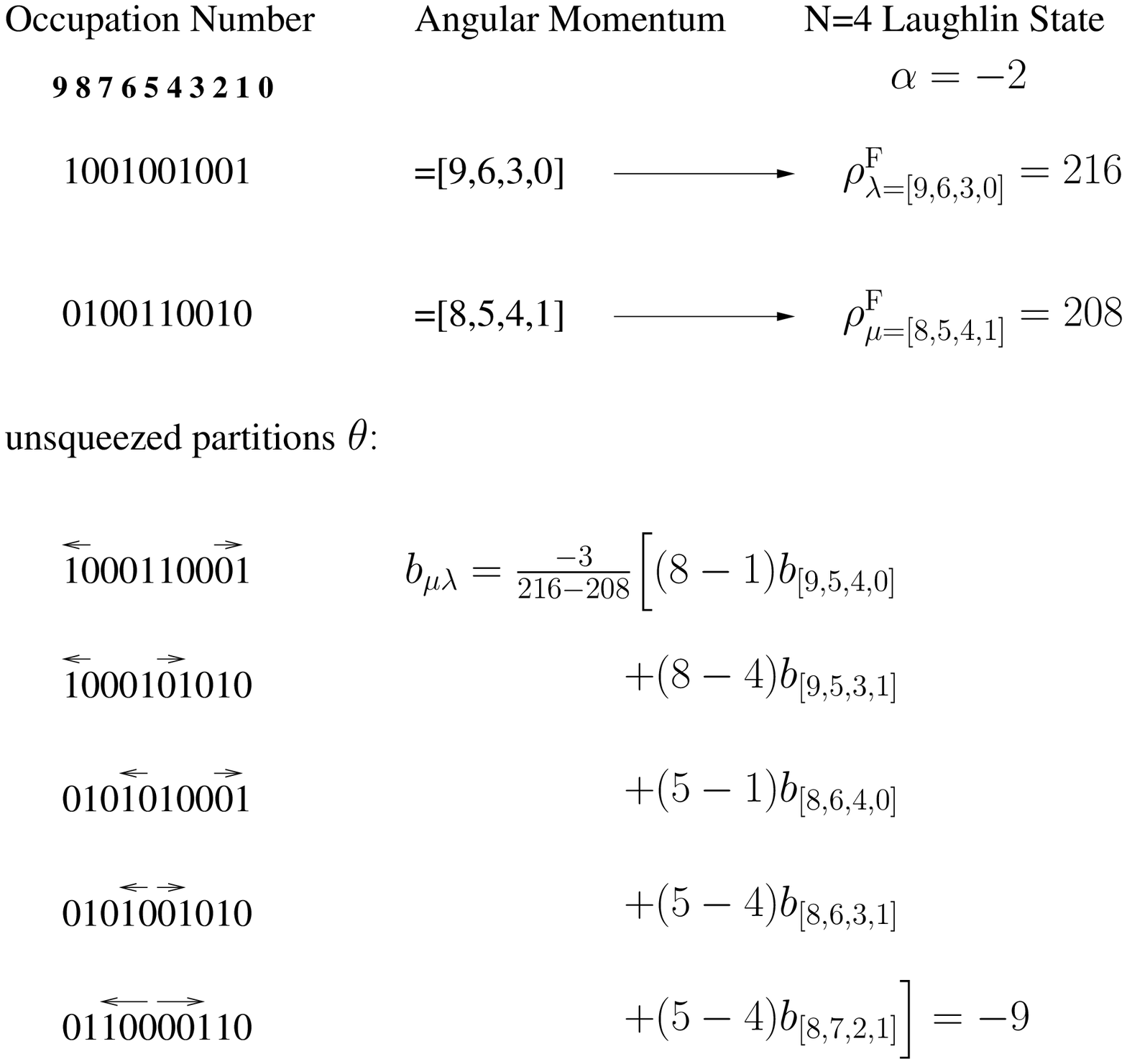}
  \end{minipage}
  \caption{The recurrence relation~\eqref{slater1} for
    the $N=4$ particle $\nu=1/3$ Laughlin state. The partitions are written in
    decreasing order of orbital angular momentum, which ranges from
    $9$ to $0$ in the case considered. The coefficient
    of the partition $\mu=0100110010$ is computed with the knowledge of
    the coefficient for the partitions dominating $\mu$.}
\label{coeffi1}
\vspace{-0pt}
\end{figure}

\noindent where $\rho^{\text{F}}_\lambda (\alpha) = \sum_i \lambda_i
(\lambda_i +2i (1-1/\alpha))$. Similar to the bosonic recurrence
formula in~\eqref{recbos}, the sum in~\eqref{slater2} extends over
all partitions $\theta = [\mu_1, \dots , \mu_i+s, \dots, \mu_j-s,
\dots, \mu_N]$ that dominate the partition $\mu=[\mu_1,\dots,\mu_N]$
and are squeezed from the root partition $\lambda$. A new factor
$(-1)N^{\text{SW}}$ appears: a sign according to the even/oddness of
the number of transpositions (swaps) of particles from a given
dominating partition $\theta$ back to $\mu$.  This term appears since
the reordering of the partition in Slater determinant language may
cause a minus sign due to the fermionic anticommutation relations.
$N_{\text{SW}}$ starts from zero for partition $\mu$ and advances by
one unit every time the momentum of the unsqueezed electron passes
through the value of the momentum for another electron. As a further
difference from the bosonic Jack recurrence relation, the terms summed
in Eq.~\ref{slater2} do not explicitly depend on the partition
$\theta$.  This is because the rescaling of the $s$ in~\eqref{slater1}
exactly cancels the term's dependence on $s$. For $\alpha=-(k+1)$,
\eqref{slater2} gives the coefficients of the fermionic Read-Rezayi
states (an example computation of the partition coefficient for the
$\nu=1/3$ Laughlin state is shown in Fig.~\ref{coeffi1}). The complete
decomposition of the $N=4$ particle $\nu=1/3$ Laughlin state is given
in Appendix~\ref{app:mono}.  As in the bosonic case, this is an
advance in the numerical computation of the coefficients: the
computational effort required to generate the state scales linearly
with basis size. This approach has already been applied to increase the maximally
reachable system in finite size studies~\cite{papic-09prb245325}.

\section{Product Rules}
\label{sec:product}


The coefficients of the monomials (Slaters) in the bosonic (fermionic)
FQH states exhibit a hidden symmetry found in
Ref.~\onlinecite{bernevig-09prl206801} and named product rule. The
product rule is valid in \emph{any quantum mechanical normalization},
be it on the plane, sphere, cylinder, disk or any other genus-$0$
geometry. It is however easiest to see and explain in the basis of
Eq.~\ref{recbos} and Eq.~\ref{slater2}, for which we have already
developed the formalism of the previous sections. Once a state is
obtained in this basis, it is only a matter of specific change in the
normalization of the free many-body wave functions to go between
different genus-$0$ geometries. The product rule found
in~\cite{bernevig-09prl206801} and explained in detail in this section
is valid not only for FQH Jacks but for \emph{all} Jacks at \emph{any}
$\alpha$, of any partition $\lambda$. Furthermore, as shown in
Section~\ref{sec:cft} below, the product rule is a property of quantum
Hall trial states even beyond Jack
polynomials~\cite{simon-10prb121301,mcs,bonderson-08prb125323}.

\begin{figure}[t]
  \begin{minipage}[l]{0.99\linewidth}
    \includegraphics[width=\linewidth]{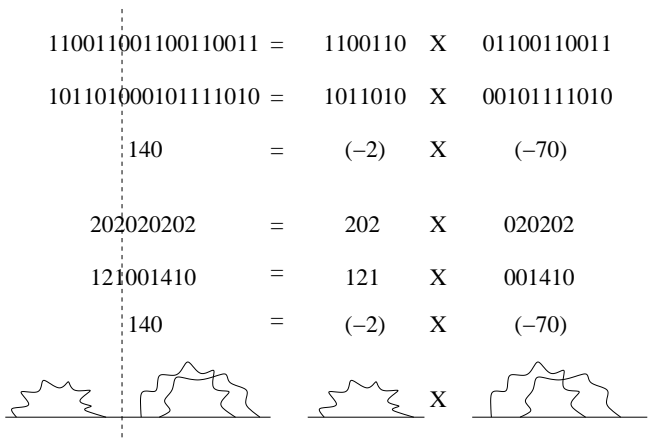}
  \end{minipage}
  \caption{Product rule applied for a partition of the $N=10$
    MR state in fermionic and bosonic notation. The given
    partition can be related to the expansion coefficients of product
    of smaller systems, in this case the $N=4$ and the $N=6$
    MR state partitions.}
\label{product1}
\vspace{-0pt}
\end{figure}

We consider a Jack state generated from~\eqref{recbos}
(or~\eqref{slater2}) that can serve as a suitable example to
demonstrate the product rule. We discuss the fermionic MR state of
$N=10$ particles. This state can be written as a linear superposition
of Slater determinants squeezed from the root partition $n(\lambda) =
110011001100110011$. We pick a configuration squeezed from $\lambda$
that has the special property that two parts of the partition can be
identified as squeezed from individual root partitions for smaller
systems sizes (see Fig.~\ref{coeffi1}). Let us consider
$101101000101111010$. We observe that the first 7 orbitals from the
left, i.e.  $1011010$, can be squeezed from the $N=4$ partition
$1100110$. The remainder right part, i.e.  $00101111010$, can be
squeezed from the $N=6$ partition $01100110011$.  We find that the
product of the two coefficients obtained from the $N=6$ part and the
disconnected $N=4$ part ($-70$ and $-2$ respectively) gives the
coefficient $(-70)\cdot (-2)=140$ of the $N=10$ partition.
The product rule (symmetry) allows the computation of a certain set of
coefficients of an $N$-particle state from the knowledge of the state
for $N-1$ particles.  This hints at similarities with Feynman
disconnected diagram summation in interacting systems, where the total
contribution is given by the product of the disconnected components
(Fig.~\ref{product1}). As we show below, the product rule
approximation succeeds in keeping the essential part of the
correlation of the FQH state.  We first prove the product rule for
fermionic Jacks by induction principle; a similar proof can be
obtained for the bosonic Jacks.

{\it Basic induction case.} Assume we start with any fermionic
polynomial $S^\alpha_\lambda(z_1,\dots,z_N)$ of a configuration $\mu
\le \lambda$ that can be divided in two disconnected sets:
$\mu=(\mu_A,\mu_B)$, with $N_A$ particles in the first subpart $A$ and
$N_B=N-N_A$ particles in the second subpart $B$.  This means that
$\mu_A$ is squeezable from an $N_A$-particle root partition
$\lambda_A=[\lambda_1,\dots, \lambda_{N_A}]$, and $\mu_B$ is squeezed
from the $N_B$-particle root partition
$\lambda_B=[\lambda_{N_A+1},\dots,\lambda_N]$. The basic induction
case is given by any partition of the monomials in
$S^\alpha_\lambda(z_1,\dots,z_N)$ for which the product rule holds.
Trivially, for this purpose we can choose the root partition
$\lambda$. By definition, it has coefficient $1$, and we can think of
it as being separated into any product of two subpart root partitions.
Again, all these have coefficient $1$ by definition, so that the
product rule holds for the root partition itself.

{\it Induction hypothesis.} We now assume the product rule is valid
for {\it all} partitions $\theta$ sharing the separable form
$\theta=(\theta_A,\theta_B)$, where $\mu_A<\theta_A \le \lambda_A$ and
$\mu_B < \theta_B \le \lambda_B$. As shown in Eq.~\ref{slater1}, the
coefficients of $S_\lambda^\alpha$ are given as a recursion from
partitions that dominate $\mu$. By construction, any partition
dominating $\mu$ and entering~\eqref{slater1} is also separable
according to $\theta=(\theta_A,\theta_B)$. As the unsqueezed operation
is a two-body operation, the sum over all dominating partitions
$\theta$ can be decomposed into $\sum_{\theta; \; \mu<\theta \le
  \lambda}=\sum_{\mu_A < \theta_A \le \lambda_A}+\sum_{\mu_B< \theta_B
  \le \lambda_B}$. In particular, the summation over the individual
partition entries $\mu_i$ only mixes $\mu_i, \mu_j$ of the left hand
side $A$ and right hand side $B$ separately, while the remainder right
(left) part remains unchanged. Partition-wise, the first sum reads
$(\theta_A,\mu_B)$, while the second reads $(\mu_A,\theta_B)$.
Finally, we assume that all partitions dominating $\mu$ satisfy the
product rule:
\begin{equation}
b_{\lambda(\theta_A,\mu_B)} (\alpha) = b_{\lambda (\theta_A, \lambda_B)}(\alpha) b_{\lambda (\lambda_A,\mu_B)}(\alpha),
\end{equation}
where $(\theta_A,\lambda_B)$ denotes the partition formed by
$\theta_A$ and the remainder root state partition of part $B$,
$\lambda_B$. This holds vice versa for $(\lambda_A,\theta_B)$.

{\it Induction proof.} Let us consider the coefficient $b_{\lambda,
  \mu}$. By induction hypothesis, we can rewrite Eq.~\ref{slater1} as (we skip the argument $\alpha$ in the notation for the
coefficients $b$):
\begin{widetext}
\begin{eqnarray}
&&b_{\lambda \mu} = \frac{2(\frac{1}{\alpha}-1)}{\rho_{\lambda}^{\text{F}}-\rho_{\mu}^{\text{F}}} \left(\sum_{\theta_A; \; \mu_A<\theta_A \le \lambda_A} (\mu_i^A-\mu_j^A) b_{\lambda (\theta_A,\mu_B)}  (-1)^{N_{\text{SW}}} + \sum_{\theta_B; \; \mu_B < \theta_B \le \lambda_B} (\mu_i^B-\mu_j^B) b_{\lambda (\mu_A, \theta_B)} (-1)^{N_{\text{SW}}} \right)\nonumber \\
&=& \frac{2(\frac{1}{\alpha}-1)}{\rho_{\lambda}^{\text{F}}-\rho_{\mu}^{\text{F}}}\left( b_{\lambda(\lambda_A,\mu_B)}\hspace{-10pt}\sum_{\theta_A; \; \mu_A<\theta_A \le \lambda_A}\hspace{-10pt} (\mu_i^A-\mu_j^A) b_{\lambda (\theta_A,\lambda_B)}  (-1)^{N_{\text{SW}}} + b_{\lambda (\mu_A, \lambda_B)} \hspace{-10pt}\sum_{\theta_B; \; \mu_B < \theta_B \le \lambda_B}\hspace{-10pt} (\mu_i^B-\mu_j^B) b_{\lambda (\lambda_A, \theta_B)} (-1)^{N_{\text{SW}}} \right). \nonumber \\
\end{eqnarray}
Writing out the coefficients $b_{\lambda(\lambda_A,\mu_B)}$ and
$b_{\lambda (\mu_A, \lambda_B)}$ according to Eq.~\ref{slater1}, we then use
$\rho_\lambda^{\text{F}}(\alpha)-\rho_\mu^{\text{F}}(\alpha)=\rho_{\lambda}^{\text{F}}(\alpha)-\rho_{\mu_A
  \lambda_B}^{\text{F}}(\alpha) +
\rho_{\lambda}^{\text{F}}(\alpha)-\rho_{\lambda_A
  \mu_B}^{\text{F}}(\alpha)$ to get
\begin{eqnarray}
b_{\lambda \mu} &=& \frac{2(\frac{1}{\alpha}-1)}{\rho_{\lambda}^{\text{F}}(\alpha)-\rho_{\mu_A \lambda_B}^{\text{F}}(\alpha) + \rho_{\lambda}^{\text{F}}(\alpha)-\rho_{\lambda_A \mu_B}^{\text{F}}(\alpha)}\left(  \frac{2(\frac{1}{\alpha}-1)}{\rho_{\lambda}^{\text{F}}(\alpha)-\rho_{\mu_A \lambda_B}^{\text{F}}(\alpha)} + \frac{2(\frac{1}{\alpha}-1)}{\rho_{\lambda}^{\text{F}}(\alpha)-\rho_{\lambda_A \mu_B}^{\text{F}}(\alpha)}\right) \nonumber \\
&& \times \sum_{\mu_A<\theta_A\le \lambda_A} \sum_{\mu_B<\theta_B\le \lambda_B} (\mu_i^A-\mu_j^A) (\mu_i^B-\mu_j^B) b_{\lambda (\lambda_A,\theta_B)} b_{\lambda (\theta_B,\lambda_A)} (-1)^{N_{\text{SW,A}}+N_{\text{SW,B}}} \nonumber \\
&=& b_{\lambda (\mu_A,\lambda_B)} \cdot b_{\lambda (\lambda_A,\mu_B)} \quad \text{q.e.d.}
\end{eqnarray}
\end{widetext}
A similar line of reasoning applies to the bosonic case. We have thus
proved the product rule symmetry for this type of piece-separable
configurations. This is valid for all bosonic and fermionic Jack
polynomials, and hence for all Read-Rezayi states. However, the
product rule even applies to a much larger range of polynomials
(Section~\ref{sec:cft}). Following similar steps as above, the product rule can
also be explicitly derived for spin-unpolarized states such as the
Haldane-Rezayi state discussed below in Section~\ref{sec:hr}.

For bosonic states, there are certain classes of partitions where it
is not immediately clear how to apply the product rule symmetry.
Let us look again at the $N=10$ particle MR state in bosonic
notation; this state is squeezed from the root partition
$\lambda=202020202$. 

{\it Partitions type I.} An easy application of the product exists for 
configurations such as $p_1=040000600$. We identify the first $4$
orbitals to be squeezed from $\lambda_A=2020$, while the remainder
orbitals are squeezed from $\lambda_B=020202$; we find the product
rule to hold (see also Fig.~\ref{product1}).

\begin{figure}[t]
  \begin{minipage}[l]{0.99\linewidth}
    \includegraphics[width=\linewidth]{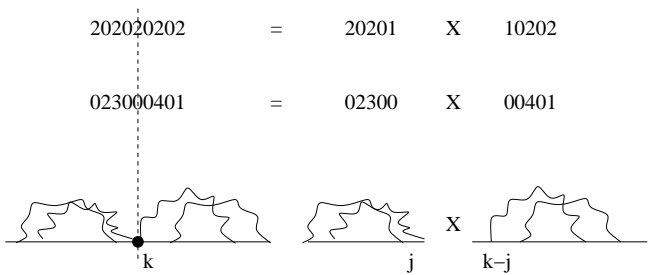}
  \end{minipage}
  \caption{Generalized product rule for certain partitions of
    $Z_k$-parafermion non-Abelian states in bosonic notation (shown: a partition of the
    $N=10$ Moore Read state). In terms of diagrammatic representation
    of interaction vertices, the orbital along the partition cut is
    assigned a label $k$ for the occupancy of the associated root
    partition orbital. The diagrammatic decomposition proceeds by
    splitting the $k$ particle into $j$ to the left and $k-j$ to the
    right. This way, the product rule holds for this class of
    configurations.}
\label{product2}
\vspace{-0pt}
\end{figure}

{\it Partitions type II.} Let us analyze the configuration
$p_2=023000401$. We can split the configuration in a $5$ particle
separation to the left and a $5$ particle separation to the right of
the cut in Fig.~\ref{product2}.  Both parts are disconnected in terms
of squeezing operations on the particles.  However, what are the root
partitions from which we generate the subparts? We have to split one
doubly occupied orbital of the associated root partition $202020202$ as in
Fig.~\ref{product2}.
We double copy this orbital and distribute the particles in both
subparts. We consider $2020{\bf 1}$ and ${\bf 1}0202$ as the root
partitions for subpart $A$ and $B$: together they make up $202020202$
but the orbital where the bold particle ${\bf 1}$ is placed is taken
to belong to both parts $A$ and $B$.  At the same time, we double copy
the $5$th orbital of $p_2$,
i.e. $0230{\bf 0}$ and ${\bf 0}0401$ (see Fig.~\ref{product2}). 
Following this recipe, we find that the product rule holds for these
type II configurations. We can trace back separable configurations of
type $p_2$ to product rule compositions of smaller system size
(Fig.~\ref{product2}).

  Let us investigate the accuracy of the approximated state for $N$
  particles built from the product rule using the exact
  states up to $N-1$ particles. To begin with, an important quantity
  is the number of monomials (or Slater determinants) whose
  configurations conform to the product rule. Tables~\ref{tab:productrulelaughlin}
  and~\ref{tab:productruleMR} show the ratio between this number and
  the total size of the squeezed Hilbert space for the Laughlin state
  ($\nu=1/2$ and $\nu=1/3$) and the MR state, respectively. As a rule
  of thumb, the product rule allows to construct more than a third and
  less than a half of the total Hilbert space. This ratio decreases
  with increasing system size, but remains a finite $> 1/3$ fraction of the Hilbert space. 

  The important question is how much of the exact state is kept in
  this part of the Hilbert space generated by the product rule. We
  compute the overlap between the exact state for $N$ particles and
  the state constructed only from the product rule. The overlap is
  taken using the scalar product of the sphere geometry. We find that,
  involving only type I partitions for the Laughlin state (Table
  \ref{tab:productrulelaughlin}) or type I and type II for the MR
  state (Table \ref{tab:productruleMR}), the state approximated by the
  product rule has $> 99.9 \%$ overlap with the exact state. This
  tells us that the monomials that are generated by the product rule
  contain almost all of the exact state by overlap despite comprising
  only $\sim 1/3$ of its Hilbert space. In all cases we consider, the
  overlap has the peculiar feature that it {\it increases} with system
  size (by contrast, any comparison between a model state and the
  ground state of some realistic interaction would exhibit the
  opposite behavior). As such, this provides indication that the
  product rule symmetry of quantum Hall trial states becomes exact in
  the thermodynamic limit. Fermionic states show a very similar
  behavior (see Table \ref{tab:productrulelaughlin} for the Laughlin
  state at $\nu=1/3$). The overlaps are also very high but not as good
  as their bosonic counterpart.  In our opinion, the product rule
  should be an essential ingredient of future DMRG studies.

  For now, we have only considered partitions subject to the product
  rule construction which can be decomposed into a pair of ground
  state partitions of smaller systems size.  In fact, one can further
  improve the procedure as we find for the specific case of Laughlin
  states. The product rule can be applied not only when considering
  disconnected squeezing sequences from the root partition, but also
  from a partition such as $10010001\|10001$ with a cut between the
  two consecutive particles in the $8$th and $9$th orbital. In this
  case, to reconstruct the Slater determinant weight, one needs to
  glue together two Laughlin states with one quasi-hole excitation each
  (which are also Jack polynomials with the same $\alpha$ parameter as
  the ground state). In this example, this corresponds to considering
  the Jacks of roots $10010001$ and $10001$. The only missing
  information we additionally need to know is the weight of a Slater
  determinant in a Laughlin state which is obtained from the root
  configuration by a single squeezing of two neighboring particles
  $...1001...$ into two consecutive occupied orbitals $...0110...$. It
  can be shown that this weight is always equal to $\-3$ in the basis
  described in Eq. (\ref{slater1}). The improvement of the overlap
  including this additional rule is shown in Table
  \ref{tab:productrulelaughlin}.
 
\begin{widetext}
\begin{table*}[t]
\begin{center}
\begin{tabular}{l|ccccccccc}
$N$ & 8 & 9 & 10 & 11 & 12 & 13 & 14 & 15 & 16 \\
\hline
\hline
$\nu=1/2$ prod. rules & 43.76\% & 42.01\%  & 40.76\% & 39.93\% & 39.52\% & 39.32\% & 39.24\% & 39.19\% & 39.16\% \\
$\nu=1/2$ overlap & 0.9933 & 0.9947 & 0.9947 & 0.9956 & 0.9963 & 0.9968 & 0.9972 & 0.9977 & 0.9979 \\
\hline
$\nu=1/3$ prod. rules & 47.68\% & 46.41\%  & 45.33\% & 44.45\% & 43.73\% & 43.11\% & 42.56\% & 42.08\% & 41.65\% \\
$\nu=1/3$ overlap & 0.9502 & 0.9534 & 0.9557 & 0.9573 & 0.9585 & 0.9593 & 0.9599 & 0.9603 & 0.9605\\
\hline
$\nu=1/3$ prod. rules + qh& 73.92\% & 72.52\% & 71.35\% & 70.40\% & 69.61\% & 68.90\% & 68.27\% & 67.70\% & 67.18\%  \\
$\nu=1/3$ overlap + qh & .9938 & 0.9944 & 0.9949 & 0.9952 & 0.9955 & 0.9956 & 0.9957 & 0.9958 & 0.9958 \\
\hline
\end{tabular}
\end{center}
\caption{Shown: Percentage of Hilbert space that can be constructed from the product rule
  and overlap between the exact full state and the state built from the
  product rule for sphere geometry. $N$ is the number of particles
  and the overlap is defined as the absolute value of the scalar
  product. The first two rows of data are the results for the
  $\nu=1/2$ bosonic Laughlin state for different numbers of particles. The third and fourth row of data
  are the results for the $\nu=1/3$ fermionic Laughlin state relying
  only on the knowledge of the Slaters determined by ground state
  partitions, while the fifth and sixth row additionally take into account information stemming from the single quasi-hole excitations.}
\label{tab:productrulelaughlin}
\vspace{-5pt}
\end{table*}
\end{widetext}

\begin{widetext}
\begin{table*}[t]
\begin{center}
\begin{tabular}{l|cccccccc}
N & 8 & 10 & 12 & 14 & 16 & 18 & 20 & 22 \\
\hline
\hline
squeezed dim.& 119 & 1070 & 10751 & 116287 & 1326581 & 15756587 & 193181910 & 2429921124\\
\hline
prod. rules (type I) & 27.73\% & 25.23\% & 23.58\% & 22.48\% & 21.63\% & 20.95\% & 20.40\% & 19.95\% \\
overlap (type I)& 0.8858 & 0.9070 & 0.9188 & 0.9262 & 0.9311 & 0.9344 & 0.9367 & 0.9383 \\
\hline
prod. rules (type I + type II) & 72.27\% & 70.09\% & 68.48\% & 66.98\% & 65.79\% & 64.82\% & 64.01\% & 63.34\%\\
overlap (type I + type II)& 0.9875 & 0.9895 & 0.9919 & 0.9930 & 0.9936 & 0.9941 & 0.9944 & 0.9946\\
\hline
\end{tabular}
\end{center}
\caption{Shown: percentage of Hilbert space that is constructed from the
  product rule and overlap between the complete MR state and the state
  built from product rule on the sphere geometry. $N$ is the number of
  particles and the overlap is defined as the absolute value of the
  scalar product. The first row of data is the total dimension of the
  squeezed Hilbert space for different system sizes. The second and
  third row are the overlap results obtained when only type I
  partitions are taken into account, while fourth and fifth row show
  the results obtained by involving both type I and type II partitions.}
\label{tab:productruleMR}
\vspace{-5pt}
\end{table*}
\end{widetext}

\section{Haldane-Rezayi state}  
\label{sec:hr}

\subsection{Basic properties}

We now turn to the generalization of the bosonic and 
fermionic states involving the spin degree of freedom of the
constituent particles. In the following, we discuss the Haldane-Rezayi
(HR) spin singlet state~\cite{haldane-88prl956}. The HR state was originally proposed as a trial state for the
incompressible plateau state at $\nu=5/2$~\cite{haldane-88prl956}.  As
opposed to the spin-polarized Moore-Read (Pfaffian) state at identical
filling~\cite{Moore-91npb362,greiter-91prl3205}, it is a spin singlet.
The degree of spin polarization is still an experimental issue that is
not yet settled in the $\nu=5/2$ state. The general belief, supported
by numerical evidence from exact diagonalization studies, is that the
Moore-Read state is the promising candidate to explain the $\nu=5/2$
state~\cite{eisenstein-88prl997,Morf98prl1505,Greiter-92ncb567,ino-99prl4902}.
The HR state attained considerable attention since it can be
interpreted as describing the transition point between the strong and
weak pairing phases of a spin-singlet $d$-wave BCS superconductor.
This physical interpretation is supported by the realization that the
bulk CFT is a non-unitary $c=-2$ theory, which is expected to have
gapless bulk excitations~\cite{gurarie-97npb513}.

The HR state is given by:
\begin{eqnarray}
\Phi_{\text{HR}} &=& \mathcal{A}_{z,w} \left( \frac{1}{(z_1-w_1)^2} \frac{1}{(z_2-w_2)^2} \dots  \right) \nonumber \\
&&\times \prod_{i<j}^N (z_i - z_j)^2 \prod_{i<j}^{N} (w_i-w_j)^2 \prod_{i,j}^N (z_i-w_j)^2,\nonumber \\
\label{eq:hrstate}
\end{eqnarray}
where $N$ is the number of spin $\uparrow$ and spin $\downarrow$
particles with positions denoted by $z$ and $w$, $\mathcal{A}_{z,w}$
is the antisymmetrizer over the $z$ particles and $w$ particles
separately. Due to the antisymmetric non-$U(1)$ prefactor in the first
line of~\eqref{eq:hrstate} and the evenness of the Jastrow factors,
the state in~\eqref{eq:hrstate} is fermionic although a bosonic
variant can also be written. Let us look at the clustering conditions at
the level of first quantization. The wave function dies as
the 2nd power of the difference between two equal spin coordinates.
However, the prefactor of the Jastrow factors removes the 2nd order
inter-spin zeroes induced by the last Jastrow factor, allowing for
configurations where one $\uparrow$ and one $\downarrow$ spin sit on
the same site.

In order to span the Hilbert space of spinful states, we again start
from the single particle Landau orbitals. However, in this case, the
single particle orbitals possess an additional spin quantum number
$\sigma$ taking on the values $\uparrow$ or $\downarrow$. The free
many-body basis is given by spinful Slater determinants
$\text{sl}_{\lambda_{A \uparrow}, \lambda_{B
    \downarrow}}=\mathcal{A}(z_1^{\lambda_{1, A \uparrow}}, \dots ,
z_N^{\lambda_{N, A\uparrow}}, w_1^{\lambda_{1, B\downarrow}}, \dots,
w_N^{\lambda_{N, B\downarrow}})$, $\lambda_{i, A\uparrow}$ and
$\lambda_{i, B\downarrow}$ label the momenta of the $i$th $\uparrow$
spin and $i$th $\downarrow$ spin particles, respectively. 

The squeezing operation is defined as before, and applies equally to
squeezing between $\uparrow$ or $\downarrow$ particles.  The root
partition of the HR state is given by $X000X \dots 0X$, where
$X=\uparrow \downarrow$, i.e.  $X$ denotes an orbital occupied by both
spins. This is consistent with the first quantized clustering
condition and with the filling $\nu=1/2$ in the highest partially
populated Landau level.  We consider the decomposition of the HR state
as Slater determinants:
\begin{equation}
\Phi_{\text{HR}}= \sum_{\lambda_A \uparrow, \lambda_B \downarrow \le X000X000X\dots} c_{\lambda_{A} \uparrow, \lambda_{B} \downarrow} \text{sl}_{\lambda_{A} \uparrow,\lambda_{ B} \downarrow},
\end{equation}
where the sum extends over all spinful partitions obtained by
squeezing operations on the root partition $X000X\dots 0X$. Let us
consider the 4-particle HR root configuration $X000X$. It can be
expressed in partition language as $[4\uparrow, 4\downarrow,
0\uparrow, 0\downarrow]$ or $[4\uparrow, 0\uparrow]\times[4\downarrow,
0 \downarrow]$. The latter is a factorization in $\uparrow$-spin and
$\downarrow$-spin partitions. It differs from the former by a minus
sign. Thus, to avoid ambiguities of global minus signs due to different
orderings of fermionic operators, we order all partition entries (i.e.
$\uparrow$ and $\downarrow$ spin) first by momenta in decreasing
order. For a given momentum, we write the $\uparrow$-spin entry before
the $\downarrow$-spin entry. Whenever a factorized partition notation
of $\uparrow$ spin and $\downarrow$ spin momenta appears in the
following text, it is only for reasons of presentation. The ordering
of fermions should always be interpreted as explained above. Moreover,
Slater determinants differing by an overall spin
rotation have equal coefficients. This is so since the HR state is a
spin singlet and thus spin rotationally invariant.



\subsection{Differential action for the HR state}
From conformal field theory considerations,  $\Phi_0(z,w)=\mathcal{A}_{z,w}
(\frac{1}{(z_1-w_1)^2}\frac{1}{(z_2-w_2)^2 \dots })$ satisfies the
following differential equation~\cite{DiFrancescoMathieuSenechal97}:
\begin{equation}
\left(\frac{1}{2} \frac{\partial^2}{\partial z_i^2} -\sum_{j\ne i} \left( \frac{1}{(z_i-w_j)^2}+\frac{1}{z_i-w_j} \frac{\partial }{\partial w_j}\right)\right) \Phi_0=0,
\label{eq:hr-nu1}
\end{equation}
with the same equation for $z \leftrightarrow w$.
Following section \ref{sec:fermion}, we use this equation to obtain an
operator for 
which $\Phi_{\text{HR}}$ is an eigenstate. We rewrite the derivatives acting on
$\Phi_0$ as derivatives acting on $\Phi_{\text{HR}}$. This amounts to
taking into account the additional derivative acting on Jastrow
factors for both spin species. Using the intermediate steps
explained in Appendix~\ref{app:hr}, we derive the following differential
equation:
\begin{widetext}
\begin{eqnarray}
&& \left[\frac{1}{2}\sum_i^N \left(z_i \frac{\partial}{\partial z_i} \right)^2+ \left(w_i \frac{\partial}{\partial w_i}\right)^2 - (3N-2) \sum_i^N \left( z_i\frac{\partial}{\partial z_i} + w_i \frac{\partial}{\partial w_i} \right)-\frac{1}{2} \sum_{i,k} \frac{z_i+w_k}{z_i-w_k} \left(z_i \frac{\partial}{\partial z_i} - w_k \frac{\partial}{\partial w_k} \right)\right.\nonumber \\
&& \left.-\frac{1}{4}\sum_{i\ne j}\left(  \frac{z_i+z_j}{z_i-z_j} \left(z_i \frac{\partial}{\partial z_i} - z_j \frac{\partial}{\partial z_j} \right)+\frac{w_i+w_j}{w_i-w_J} \left(w_i\frac{\partial}{\partial w_i} -w_j \frac{\partial}{\partial w_j} \right) \right) + 4N(2N^2-3N+1)\right] \Phi_{\text{HR}}=0. \label{eq-hr}
\end{eqnarray}
\end{widetext}
The above equation contains only 2-body interactions. The interaction
terms are symmetric with respect to $\uparrow$-spin and
$\downarrow$-spin variables. Both the inter and intraspin interaction are
of Laplace-Beltrami type and are familiar from our previous calculations of the polarized fermionic
states.
\subsubsection{Equal spin action}
Let us first compute the action of the terms in~\eqref{eq-hr}
consisting of equal spin interactions. Once solved for one species,
e.g. the $\uparrow$ spin variables $z$, this also applies for the
$\downarrow$ spin terms. This part of the HR operator is given by:
\begin{equation}
\frac{z_i+z_j}{z_i-z_j}\left(z_i \frac{\partial}{\partial z_i} - z_j\frac{\partial}{\partial z_j} \right),
\label{equalspin}
\end{equation}
which we previously encountered as part of the fermionic LB operator.
However, the term $\sim 1/(z_i -z_j)^2$ of the LB operator is missing.
This is an important difference. It implies that the operator in Eq.~\ref{equalspin} does not
map a single \emph{general} Slater determinant of arbitrary degree
into another Slater determinant. The equivalent of~\eqref{slater1}
cannot be written in the current case by using only the operator in
Eq.~\ref{equalspin}: Acting with the operator in Eq.~\ref{equalspin}
on a single Slater determinant usually leads to a fraction.
However, the special sum of Slater determinants that comprise
the HR state \emph{does} map back into a sum of Slater determinants.
This is so since the HR state has two Jastrow terms in equal spins;
they cancel any fractions that might appear upon the action
of~\eqref{equalspin} on a Slater (see Appendix~\ref{app:hr}). A single
Jastrow factor would not.  Hence a mapping of a linear
superposition of Slaters constrained in this way maps, under action
of~\eqref{equalspin}, back to the space of Slaters.  Details of this
are given in Appendix~\ref{app:hr}. We hence force the identity:
\begin{equation}
\sum_{i \ne j}\frac{z_i+z_j}{z_i-z_j}\left(z_i \frac{\partial}{\partial z_i} - z_j \frac{\partial}{\partial z_j} \right)\sum_{\mu}a_{\mu}\text{sl}_\mu=\sum_{\mu}b_{\mu}\text{sl}_{\mu},
\label{equalspin2}
\end{equation}
where $\text{sl}_\mu$ is defined as the Slater determinant of one spin
species. The partition of the other spin species is omitted in typing
but should be implicated. Since
from the expression of the HR state we know that the linear
combination of states that form the HR state have a double zero $\sum_\mu a_\mu
\text{sl}_\mu (z_1 \dots z_M) \sim \prod_{i,j} (z_i-z_j)^2$, the
forced identity~\eqref{equalspin2} is clearly true. The coefficients
$b_\mu$ are then given in terms of the coefficients $a_\mu$. We find
\begin{eqnarray}
\sum_{\mu \le \lambda}b_{\mu} \text{sl}_{\mu}&=&\sum_{\mu \le \lambda}\sum_{i<j}(\mu_i-\mu_j)a_\mu \text{sl}_\mu \nonumber \\ &&+ 2 a_\mu (\mu_i-\mu_j) \sum_{\theta < \mu}\text{sl}_{\theta} (-1)^{N_{\text{SW}}},
\label{equal-big}
\end{eqnarray} 
where $\theta=[\mu_1, \dots, \mu_i-s, \mu_{i+1}, \dots, \mu_j+s,
\dots, \mu_N]$, and $N_{\text{SW}}$ again denotes the number of swaps
needed to reorder the partition.

\subsubsection{Different spin action}

In order to compute the term in~\eqref{eq-hr} involving action on both
spin species, we consider one single spinful Slater determinant
$\text{sl}_{\lambda_A \uparrow, \lambda_B \downarrow}$ as defined
before. Following the calculation detailed in Appendix~\ref{app:hr},
the spin-rotated partition $\text{sl}_{\lambda_B \uparrow, \lambda_A
  \downarrow}$ always appears with the same coefficient. In this way,
we find that the action of the $\uparrow \leftrightarrow \downarrow$
part of the operator also maps back to Slaters:
\begin{widetext}
\begin{eqnarray}
&&\sum_{i,k} \frac{z_i+w_k}{z_i-w_k}\left( z_i \frac{\partial}{\partial z_i} - w_k \frac{\partial}{\partial z_k} \right)\text{sl}_{\lambda_A \uparrow, \lambda_B \downarrow}\nonumber \\
&=& \sum_{i,j,\sigma; \lambda_{i,\sigma}>\lambda_{j,\bar{\sigma}}} (\lambda_{i,\sigma}-\lambda_{i, \bar{\sigma}}) \text{sl}_{\lambda_A \uparrow, \lambda_B \downarrow}
+\hspace{-10pt}\sum_{\theta_A, \theta_B; i,j,\sigma; \lambda_{i,\sigma}>\lambda_{j,\bar{\sigma}}}\hspace{-10pt} (\lambda_{i,\sigma}-\lambda_{i, \bar{\sigma}} ) \text{sl}_{\theta_A \uparrow, \theta_B \downarrow} (2-\delta_{\theta_A, \theta_B} \delta_{\lambda_{i, \sigma}-s, \lambda_{j, \bar{\sigma}}+s}) \cdot (-1)^{N_{\text{SW}\uparrow}+N_{\text{SW}\downarrow}}, \nonumber \\
\label{eq-unequal}
\end{eqnarray}
\end{widetext}
where $\theta_A, \theta_B$ are the two partitions of spin $\uparrow$
and $\downarrow$, respectively. They are obtained by squeezing only
{\it opposite} spins in $\lambda_{A \uparrow}, \lambda_{B
  \downarrow}$; $s$ parametrizes the changed partition component
obtained from squeezing spin $\uparrow$ with spin $\downarrow$:
$\theta_A=[\lambda_{1, A},\dots,\lambda_{i, A}+s,\dots,\lambda_{N,
  A}]$ and $\theta_B=[\lambda_{1, B},\dots,\lambda_{j,
  B}-s,\dots,\lambda_{N, B}]$. Thus, the summation takes each possible
pairwise combination of spin $\uparrow$ and spin $\downarrow$
particles, and squeezes them. $N_{\text{SW}\sigma}$ denotes the
reordering swaps of spins from species $\sigma$ upon this inter-spin
squeezing operation.

\subsection{Recurrence relation}

We are now ready to compute the full action of the operator in
Eq.~\eqref{eq-hr} on the Haldane-Rezayi state. The remainder terms are
of non-interacting kinetic type and straightforward to compute. By
power counting, we find:
\begin{equation}
\sum_i^{N}\left( z_i \frac{\partial}{\partial z_i} + w_i \frac{\partial}{\partial w_i}\right) \Phi_{HR}= 4N(N-1) \Phi_{HR}.
\end{equation}
The second order derivative terms give 
\begin{equation}
\sum_i \left( z_i \frac{\partial}{\partial z_i} \right)^2 \Phi_{\text{HR}}=\sum_i \lambda_i^2 \Phi_{\text{HR}}.
\end{equation}
In the intra-spin term, the sums over differences of
$\lambda$'s can be computed as
\begin{equation}
\sum_{i<j}^N (\lambda_i - \lambda_j) = \sum_{i}^{N} (N+1-2 i )\lambda_i,
\end{equation}
while there is no similar closed form expression for the inter-spin
term.  Summing up all terms of~\eqref{eq-hr}, we find the following
equation:
\begin{widetext}
\begin{eqnarray}
&& \sum_{\lambda_A \uparrow, \lambda_B \downarrow \le X000X000X\dots} c_{\lambda_{A} \uparrow, \lambda_{B} \downarrow} \text{sl}_{\lambda_{A} \uparrow,\lambda_{ B} \downarrow}\left(2N(3N^2-4N+1) - \frac{1}{2}\sum_i (\lambda_{A i}^2 + \lambda_{B i}^2 + 2 i \lambda_{A i } + 2 i \lambda_{B i}) \right. \nonumber \\
&& \left. + \frac{1}{2}\sum_{i,j,\sigma; \lambda_{i, \sigma} > \lambda_{j,\bar{\sigma}}} (\lambda_{i, \sigma}-\lambda_{i, \bar{\sigma}}) \right) \nonumber \\
&=& -\sum_{\lambda_A \uparrow, \lambda_B \downarrow \le X000X000X\dots} c_{\lambda_a \uparrow, \lambda_B \downarrow} \left( \sum_{\theta_A < \lambda_A} (\lambda_{A,i}-\lambda_{A,j}) \text{sl}_{\theta_A \uparrow, \lambda_B \downarrow} (-1)^{N_{\text{sw}}} + \sum_{\theta_B< \lambda_B} (\lambda_{B,i}-\lambda_{B,j}) \text{sl}_{\lambda_A \uparrow, \theta_B \downarrow} (-1)^{N_{\text{SW}}}\right.\nonumber \\
&&\left. + \frac{1}{2} \sum_{i,j,\sigma; \lambda_{i,\sigma}> \lambda_j,\bar{\sigma}} (\lambda_{i,\sigma} -\lambda_{i,\bar{\sigma}}) \text{sl}_{\theta_{A,\uparrow}, \theta_{B,\downarrow}} (2 - \delta_{\theta_A, \theta_B} \delta_{\lambda_{i,\sigma}-s, \lambda_{j,\bar{\sigma}}+s}) (-1)^{N_{\text{SW},\uparrow}+N_{\text{SW},\downarrow}} \right).
\label{hraction}
\end{eqnarray}
In Eq.~\ref{hraction}, we have grouped all diagonal terms to the left
and all interactions to the right. Accordingly, the right hand side is
made of three parts: parts of sums of spin-$\uparrow$ partitions
squeezed from the spin-$\uparrow$ partition $\lambda_A$, parts of sums
of spin-$\downarrow$ partitions squeezed from the spin-$\downarrow$
partition $\lambda_B$, and parts of sums of spin-$\uparrow$ and
spin-$\downarrow$ partitions squeezed only from particles of different
spin from the partition $[\lambda_A \uparrow, \lambda_B \downarrow]$.
We now define the quantities
\begin{equation}
\rho_{\lambda_A \uparrow, \lambda_B \downarrow} = \frac{1}{2} \sum_i \left( \lambda_{A,i}^2 + \lambda_{B,i}^2 + 2 i \lambda_{A,i} + 2 i \lambda_{B,i} \right) -\frac{1}{2} \sum_{i,j,\sigma; \lambda_{i,\sigma}>\lambda_{j,\bar{\sigma}}} (\lambda_{i,\sigma} -\lambda_{i, \bar{\sigma}}).
\end{equation}
We can then immediately show that the root partition obeys
$\rho_{X000X \dots X}= 2N(3N^2-4N+1)$, and
thus corresponds to the constant terms on the left side
of~\eqref{hraction}. Now, equating the coefficient of every Slater
determinant, we derive a recurrence relation for the
coefficients of the decomposition:

\begin{eqnarray}
&&c_{\mu_A \uparrow, \mu_B \downarrow}= \frac{1}{\rho_{X000X000X\dots}-\rho_{\mu_A \uparrow, \mu_B \downarrow}} \nonumber \\
&&\left( \sum_{\theta_A; \mu_A < \theta_A \le \uparrow 000 \uparrow 000 \uparrow \dots \;\; \theta_A = [\mu_{A,1}, \dots, \mu_{A,i}+t, \dots, \mu_{A,j}-t,\dots, \mu_{A,N}]} (\mu_{A,i}-\mu_{A,j}+2t) c_{\theta_{A}\uparrow, \mu_{B} \downarrow} \cdot (-1)^{N_{\text{SW}}} \right. \nonumber \\
&& + \sum_{\theta_B; \mu_B < \theta_B \le \downarrow 000 \downarrow 000 \downarrow \dots  \;\; \theta_B = [\mu_{B,1},\dots, \mu_{B,i}+t, \dots, \mu_{B,j}-t, \dots, \mu_{B,N}]} (\mu_{B,i} -\mu_{B,j}+2t) c_{\mu_{A} \uparrow, \theta_{B} \downarrow} \cdot (-1)^{N_{\text{SW}}} \nonumber \\
&& \left. + \frac{1}{2}\sum_{\theta_A, \theta_B; \;\; [\mu_A \uparrow, \mu_B \downarrow]<[\theta_A, \theta_B] \le [X000X000X \dots]} (\mu_{\sigma} -\mu_{\bar{\sigma}}+2t) c_{\theta_A \uparrow, \theta_B \downarrow} (2-\delta_{\mu_A, \mu_B} \delta_{\mu_i \uparrow, \mu_j \downarrow}) \cdot (-1)^{N_{\text{SW}\uparrow}+N_{\text{SW}\downarrow}} \right),
\label{rechr}
\end{eqnarray}
\noindent where the last sum over the inter-spin terms extends over
all partitions $\theta_A, \theta_B$ with the property
$[\theta_A,\theta_B]=[[\mu_{A,1}, \mu_{A,2}, \dots, \mu_{A,i}+t,
\dots, \mu_{A,N}], [\mu_{B,1}, \mu_{B,2},\dots, \mu_{B,j}-t, \dots,
\mu_{B,N}]]$ with $\mu_{A,i} \ge \mu_{B,j}$, or
$[\theta_A,\theta_B]=[[\mu_{A,1}, \mu_{A,2}, \dots, \mu_{A,j}-t,
\dots, \mu_{A,N}], [\mu_{B,1}, \mu_{B,2},\dots, \mu_{B,i}+t, \dots,
\mu_{B,N}]]$ with $\mu_{B,i} \ge \mu_{A,j}$. One explicit example of
computation is presented in Fig.~\ref{coeffi2}, and a complete
4-particle HR Slater decomposition is shown in
Appendix~\ref{app:sldec}.
\begin{figure}[t]
  \begin{minipage}[l]{0.85\linewidth}
    \includegraphics[width=\linewidth]{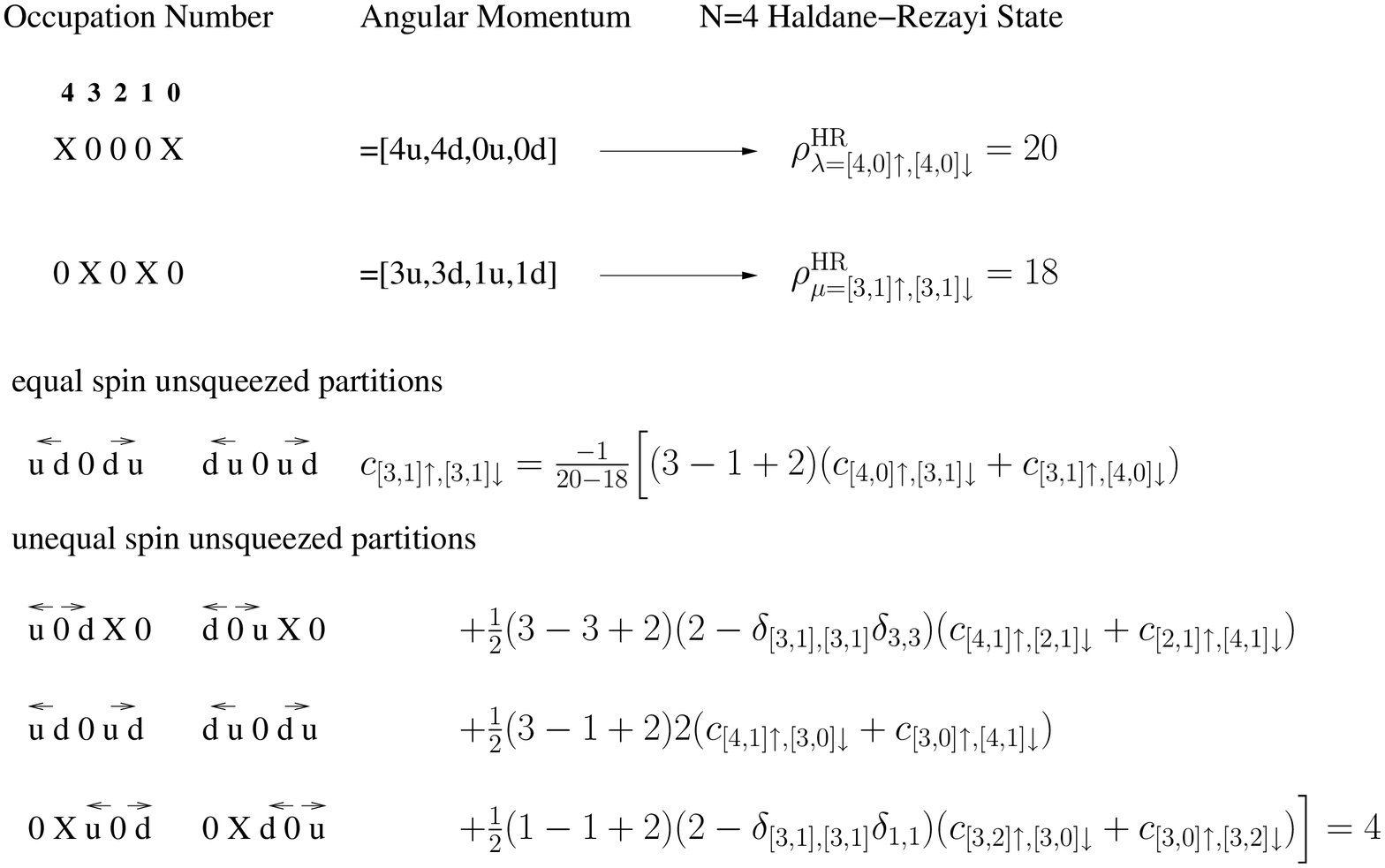}
  \end{minipage}
  \caption{Explication of the recurrence relation~\eqref{rechr} for
    the $N=4$ Haldane-Rezayi state. We have chosen to present the
    explicit computation of the coefficient of partition $\mu=0X0X0$,
    which involves all
    conceptional different terms appearing in~\eqref{rechr}.}
\label{coeffi2}
\vspace{-0pt}
\end{figure}
\end{widetext}
In analogy to~\eqref{recbos} and~\eqref{slater1}, the Slater
decomposition coefficients of the HR state can be read off
from~\eqref{rechr}. The HR state can be generated with linear
effort in Hilbert space dimension.

\subsection{Zero-weight partitions}

The recurrence relations we encountered for the spin polarized bosonic
(fermionic) states in Eq.~\ref{recbos} (Eq.~\ref{slater1}) only
produced accidental (about $1/N !$ of the total Hilbert
space dimension) zero weights for squeezed monomial (Slater)
configurations. By contrast, we observe an extensive number of zero
weight configurations from the HR recurrence relation that hints at
further structure in the HR coefficients beyond the spinful squeezed
Slater basis. The key observation is that the HR state can be written
as~\cite{milovanovic-09prb155324}
\begin{equation}
\Phi_{\text{HR}}=\prod_{i<j} (z_i-z_j)^2 (w_i-w_j)^2 P_1(z_i,w_j),
\label{p1}
\end{equation} 
where $P_1$ is an angular momentum $L=0$ polynomial squeezed from the
root partition $X 0 X 0 X 0 \dots X 0 X$. 

Interpreted in this way, the spinful $2N-$particle HR partitions are
generated from the space of partitions of $P_1$ squeezed from $X 0 X 0
X \dots X 0 X$ times partitions from the $N$-particle bosonic
$\nu=1/2$ Laughlin factor squeezed from $1010101\dots 101$ for
$\uparrow$ spins and $\downarrow$ spins, respectively.

We investigate the $2N$ particle HR partition $X000X000X \dots$, where we
count the momentum in terms of polynomial powers beginning from $m=0$
at the north pole extending to $m=4(N-1)$ at the south pole.
We define the momentum imbalance between total
$\uparrow$ spin and $\downarrow$ spin particle momentum within a
partition $\lambda$: $\vert \sum_{i} \lambda_{i, \uparrow} - \sum_{i}
\lambda_{i, \downarrow} \vert$.  
The total momentum summed over $\uparrow$ and $\downarrow$ spins,
i.e. $\sum_{i,\sigma} \lambda_{i,\sigma}$, is $4N(N-1)$.  For $N=2$,
the partitions of maximum momentum imbalance are given by $\downarrow
0 \; X \; 0 \uparrow$ and $\uparrow 0 \; X \; 0 \downarrow$ (both with
momentum imbalance $4+2-(2+0)=4$).  They are obtained by just
squeezing $\uparrow$ spins to the left and $\downarrow$ spins to the
right, and vice versa. If we arrange all $\uparrow$ and $\downarrow$
spins to generate the highest momentum imbalance between the spin
species (which means to arrange all $\uparrow$ and $\downarrow$ spins
around opposite poles), we find a partition structure $\downarrow 0
\downarrow 0 \dots \downarrow 0 X 0 \uparrow 0 \uparrow \dots 0
\uparrow$. Thus, we find a maximum imbalance of $2 N(N-1)$ for
partitions squeezed from the HR root partition.

Let us now construct the HR state starting from $P_1$. Multiplying a
$P_1$ partition of a given momentum imbalance with a $\nu=1/2$
Laughlin partition for $\uparrow$ and $\downarrow$ spins does not
change the momentum imbalance, as the total momentum added to both
spin species is equal. Thus, the available range of possible momentum
imbalance in HR partitions is given by the partitions squeezed from
$X0X0X0\dots X0X$. There, arranging the different spin species to
opposite poles gives a partition structure $\downarrow \; \downarrow
\dots \downarrow X \uparrow \; \uparrow \; \dots \uparrow$. The
maximum momentum imbalance is $N(N-1)$, and thus only one half of the
maximum imbalance for partitions squeezed from the HR root partition.
As a consequence, we can remove all squeezed partitions  in the HR state with momentum
imbalance $>N(N-1)$, as they must have zero weight. For $N=2$, this applies for the partition $\downarrow 0\; X \; 0
\uparrow$ and its spin-rotated counterpart $\uparrow 0 \; X \; 0
\downarrow$, as the momentum imbalance is $4$, while the maximum allowed
momentum imbalance is $2$ (see Appendix~\ref{app:sldec}).  While this
rule is easily implemented numerically, there are even more $0$ weight
partitions squeezed from the HR root partition, which cannot be
written in a similar closed form as the momentum imbalance constraint.
Rather, we have to explicitly check whether a partition squeezed from
the HR root partition can be generated from $P_1$ times Laughlin
partitions.  We have sketched the algorithm in
Tab.~\ref{tab:howtozero} by which partitions squeezed from the HR
root partition can be efficiently denied or accepted.

\subsection{Spinful Entanglement spectra}
\begin{widetext}
\begin{table*}[t]
\begin{center}
\begin{tabular}{cc}
  \hline\hline
  &Algorithmic steps to accept / deny a $2N-$particle HR-squeezed partition $\mu$\\
  \hline
  & For a given partition $\mu$\\
  (i) &  Check whether total momentum imbalance $\vert \sum_{i} \mu_{i, \uparrow} - \sum_{i}
  \mu_{i, \downarrow} \vert$ is $\le N (N-1)$.\\
  (ii)&  Loop over all $\uparrow$ and $\downarrow$ $N-$ particle Laughlin partitions $\kappa$ \\
  (ii.1)&  Divide partition $\mu$ by $\kappa$: Check whether all momenta are still in allowed range and no equal spin momenta are mapped\\
  & onto each other (fermionic state)  \\
  (ii.2)&  Check whether the resulting partition can be squeezed from $P_1$\\

\hline\hline 
\end{tabular}
\end{center}
\caption{Sketched algorithm to remove zero-weight partitions squeezed
  from the HR root partition $X000X000X\dots$. We apply the structure
  of the HR state according to~\eqref{p1}. By this procedure, the
  total Hilbert space is reduced to about $83\%$ of the full squeezed Hilbert space.}
\label{tab:howtozero}
\vspace{-5pt}
\end{table*}
\end{widetext}

\begin{figure}[t]
  \begin{minipage}[l]{0.99\linewidth}
    \includegraphics[width=\linewidth]{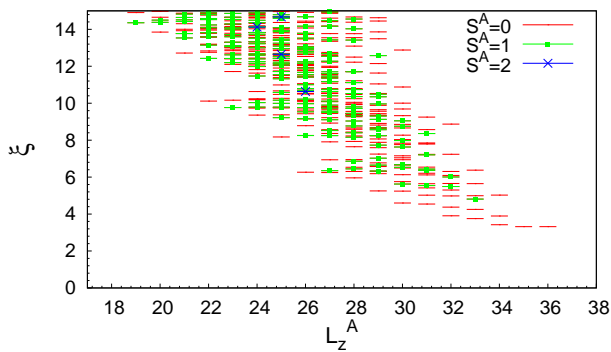}
  \end{minipage}
  \caption{(Color online) Entanglement spectrum for the $N=12$ Haldane-Rezayi state,
    half sphere cut, $N_A=6$, $S_z^A=0$ sector, entanglement levels
    plotted versus the angular momentum $L_z^A$. Levels relating to different
    spin multiplets, denoted by $S^A$,  are given in red (singlets), crossed green
    (triplets), and square blue (quintuplets). A large part of the
    main entanglement weight resides in the singlet sector. }
\label{hrspec}
\vspace{-0pt}
\end{figure}
%
We now focus our attention on the entanglement spectrum, a quantity
recently used to identify the topological fingerprint of quantum Hall
states~\cite{li-08prl010504,regnault-09prl016801,conflimit,lauch,lauch2,bergholtz,partent,schliemannarx1008,rodriguezarx1007,papicarx1008,kargarian-10prb085106,hermannsarx1009},
superconductors~\cite{noah-arx}, and topological
insulators~\cite{prodan-10prl115501,fidkowski09cm0909,turner-09cm0909},
and to detect non-local order in spin
systems~\cite{spin12,poilblanc,turnerarx1008,fidkowskiarx1008,yao-10prl080501,pollmann-09cm0910}
and Bose-Einstein condensates~\cite{liuarx1007}. We cut the sphere
into two parts and separate the orbitals into regions $A$ and
$B$. From there, we compute the reduced density matrix of region $A$
as the partial trace of $\rho$ over the degrees of freedom in region
$B$, i.e.  $\rho_A=\text{Tr}_B\rho$. We define entanglement levels
$\xi$ as the spectrum of an entanglement Hamiltonian $H_A$ that
relates to the reduced density matrix by $\rho_A=e^{-H_A}$. Depending
on the basis for which the partial trace is defined, the entanglement
levels possess certain quantum numbers associated with commuting
operators $[\rho_A,Q_A]=0$. For the HR state, the levels are specified
by quantum numbers for the number of particles, angular momentum, and
spin multiplet in $A$, i.e. $Q_A=N_A, \; L_z^A, \; S_z^A$. As the full
HR state is a spin singlet, it follows that the reduced density matrix
not only commutes with $S_z^A$, but also ${S^A}^2$ such that $\rho_A$
decomposes into spin multiplets. The $N=12$ HR spectrum for the half
sphere cut with half the particles in region $A$ is shown in
Fig.~\ref{hrspec}. In case of an even number of sites in region $A$
which gives a decomposition into integer spin multiplets, we always
choose to look at the $S_z^A=0$ spin sector where we can study all
total spin multiplets in one sector. Analyzing the single highest
angular momentum entanglement level which also is a spin singlet (see
$L_z^A=36$ in Fig.~\ref{hrspec}), we find that the eigenstate of this
entanglement level is exactly the $N=6$ HR state.  This demonstrates
the product rule property of the HR state (the fact that the highest
angular momentum entanglement level of a trial state corresponds to
the trial state itself for the number of particles in the remainder
region $A$ is found for all trial states where the product rule
applies). The product rule can also be obtained from the explicit
recurrence relation between the coefficients for the HR state
(Eq.~\ref{rechr}) by performing similar steps to the ones used in
proving the product rule for spin-polarized states. The second highest
weight sector has a single entanglement level that is degenerate in
entanglement energy with the highest momentum state. Again, for a
fermionic slater determinant description of quantum Hall trial states,
this degeneracy is constrained by symmetry of how to arrange the
slater with one less fermionic angular
momentum~\cite{bernevig-09prl206801}, and hence can again be
understood from the Slater determinant structure of the HR state. As
shown in Fig.~\ref{hrspec}, we find that both various different
angular momentum sectors and spin sectors significantly contribute to
the dominant entanglement levels of the HR state. A significant part
of entanglement of the HR state is encoded in the spin degrees of
freedom: the HR state is a spin singlet that cannot decomposed into a
product of local singlets.
\\
\section{Conformal field theory}
\label{sec:cft}

We investigate the product rule property of FQH states from the
perspective of conformal field theory (CFT). 
For the FQHE, we are concerned with chiral CFTs in 2 dimensions: all
the fields are holomorphic, \emph{i.e.} they depend on $z=x+iy$, and
not on $\bar{z}$. This is an illustration of broken time-reversal
symmetry in this context. For an in-depth introduction to CFT, we
engage the reader to Ref.~\onlinecite{difrancesco}. In this Section we will
present the key notions that are needed to derive the product rule of
quantum Hall wave functions from CFT.

The conformal dimensions are quantities that appear in the two-point
correlation functions of the fields $\Phi_i$ present in the
CFT. Conformal invariance fixes any two point function to be of the
form
 \begin{equation}
\langle \Phi_1(z_1) \Phi_2(z_2) \rangle = \frac{C_{1,2}}{(z_1-z_2)^{\Delta_1+\Delta_2}}
\end{equation}
and its scaling properties are described by the so-called conformal
dimension $\Delta_i$.

Primary fields are of special importance in a CFT. Among the infinite
number of fields of a given CFT, primary fields can be thought of as
highest weight fields.  All non-primary fields ("descendants") are
created by acting on primary fields with certain types of lowering
operators - the (negative) Virasoro modes. The two point function of
primary fields $\Phi_i$ and $\Phi_j$ reads:
\begin{equation}
\langle \Phi_i(z_1) \Phi_j(z_2) \rangle = \frac{\delta_{i,\bar{j}}}{(z_1-z_2)^{2\Delta_i}}
\end{equation}
It is non vanishing only if $\Phi_j$ and $\Phi_j$ are conjugate from
one another, namely if the identity field $1$ is produced in their
fusion. The identity is a very special field, and for our purposes it
will be the only primary field with a vanishing conformal
dimension. As a consequence it is its own conjugate, $\bar{1} = 1$. In
some cases, such as the logarithmic $c=-2$ ghost system corresponding
to the Haldane-Rezayi state encountered in Section~\ref{sec:hr}, this is not the case, and one has to
treat this cases more carefully as discussed below.

Conformal invariance also fixes the form of the correlation function
between 3 primary fields:
 \begin{eqnarray}
\lefteqn{ \langle \Phi_1(z_1) \Phi_2(z_2)\Phi_3(z_3) \rangle =  } \nonumber \\
& &  \frac{D_{1,2,3}}{(z_1-z_2)^{\Delta - 2\Delta_3}(z_2-z_3)^{\Delta - 2\Delta_1}(z_1-z_3)^{\Delta - 2\Delta_2}}  
\end{eqnarray}
where $\Delta = \Delta_1+\Delta_2+\Delta_3$. When the fusion
coefficient $D_{1,2,3}$ is non-zero, the field $\Phi_{\bar{3}}$ is
said to be produced in the fusion $\Phi_1 \times \Phi_2$. The fusion
rules encode which primary fields are produced in any fusion. For
instance a fusion rule of the form
\begin{equation}
\Phi_1 \times \Phi_2 = \Phi_a+ \Phi_b + \Phi_c
\end{equation}
tells us that the fusion of $\Phi_1$ and $\Phi_2$ only produces three
primaries: $\Phi_a,\Phi_b$ and $\Phi_c$. Implicitly, all their
descendants will also appear.

In order to compute correlation functions, fusion rules are not
sufficient. An operator product expansion (OPE) is a refinement of the
fusion rules. It is a formal (and exact) expansion of the product of
two fields, and is an implicit way to define any correlation functions
of a CFT by iteratively reducing the number of fields involved.  It
has the following generic form:
\begin{equation}
\Phi_1(z_1) \Phi_2(z_2) = \sum_{k} \frac{C_{1,2}^k}{(z_1-z_2)^{\Delta_1 + \Delta_2 - \Delta_k}} \Phi_{k}(z_2),  
\end{equation}
where the sum is formally over all fields of the CFT. Thanks to
conformal invariance, it is sufficient to know only the singular terms
- those for which $\Delta_i + \Delta_j - \Delta_k >0$ - and there will
be a finite number of them for any rational CFT. $N$-point OPEs are a
straightforward generalization:
\begin{equation}
\Phi_1(z_1)  \cdots \Phi_N(z_N) = \sum_{k} F_k(z_1,\cdots, z_N) \Phi_{k}(0) . \label{N_points OPE}
\end{equation}

We are now in a position to derive the key property of CFT correlation
functions that is responsible for the product rule.  Consider a
generic $N=N_A+N_B$-point correlation function of primary fields. For
simplicity we denote the complex coordinates of the
fields by $x_i$ and $y_j$:
 \begin{eqnarray}
\lefteqn{ C(x_i |  y_j) = } \nonumber \\
& & \langle \Phi_1(x_1) \cdots \Phi_{N_A}(x_{N_A}) \Phi_{N_A+1}( y_1)\cdots
\Phi_{N}(y_{N_B}) \rangle . \label{corr_function}
\end{eqnarray}
A particle cut between $N_A$ and $N_B$ particles is directly
related to the asymptotic behavior of the correlation function as we
spatially separate the two sets of variables $x_i$ and $y_j$. 
We consider the limit $\gamma \to \infty$ of $C(x_i | \gamma
y_j)$. Using the $N_A$ points OPE \eqref{N_points OPE} for $\Phi_1(x_1)
\cdots \Phi_{N_A}(x_{N_A})$ around $0$ we obtain
 \begin{eqnarray}
\lefteqn{ C(x_i | \gamma y_j) = } \nonumber \\
& & \sum_k F_k(x_i) \langle \Phi_{k}(0) \Phi_{N_A+1}(\gamma y_1)\cdots
\Phi_{N}(\gamma y_{N_B}) \rangle .
\end{eqnarray}
Thus conformal invariance tells us that the correlation function $ \langle
\Phi_{k}(0) \Phi_{N_A+1}(\gamma y_1)\cdots \Phi_{N}(\gamma y_{N_B})
\rangle$ - if it is non-vanishing - scales as $\gamma^{-\Delta_{N_A+1} -
  \cdots -\Delta_{N}-\Delta_k}$ as $\gamma \to \infty$.  From this
we can immediately infer that the behavior is dominated by the
(primary) field $\Phi_a$ (and its conjugate $\Phi_{\bar a}$), which is
the field with lowest conformal dimension $\Delta_a$ appearing in both
fusions:
\begin{eqnarray}
&\Phi_1 \times \cdots \times \Phi_{N_A}  \to  \Phi_a&  \\
& \Phi_{N_A+1} \times \cdots \times \Phi_{N}  \to  \Phi_{\bar a}. &
\end{eqnarray}
In addition, for a primary field $\Phi_a$ the function $F_a$ appearing in
the $N_A$ points OPE \eqref{N_points OPE} is given by the correlation function
\begin{eqnarray}
\lefteqn{  F_a(x_1,\cdots, x_{N_A}) =  \langle \Phi_1(x_1) \cdots \Phi_{N_A}(x_{N_A}) \Phi_{\bar a}(\infty) \rangle }  \nonumber \\
& & \lim_{x \to \infty} x^{2\Delta_a} \langle \Phi_1(x_1) \cdots \Phi_{N_A}(x_{N_A}) \Phi_{\bar a}(x) \rangle .
\end{eqnarray}
Finally we obtain the asymptotic behavior for
$C(x_i | \gamma y_j) $ as $\gamma \to \infty$ given by
 \begin{eqnarray}
\lefteqn{  \gamma^{\Delta_{N_A+1} + \cdots + \Delta_{N}+ \Delta_a} C(x_i | \gamma y_j) \sim } \nonumber \\
 & &  \langle \Phi_1(x_1) \cdots \Phi_{N_A}(x_{N_A}) \Phi_{\bar a}(\infty) \rangle \times \nonumber \\
 & & \langle \Phi_{a}(0) \Phi_{N_A+1}(y_1)\cdots \Phi_{N}( y_{N_B}) \rangle  \label{generic_factorization}
\end{eqnarray}
for any correlation function of primary fields $C(x_i|y_j)$
\eqref{corr_function}. Using first principles of CFT, we hence have
derived a factorized form of the correlation function spatially
separating two sets of variables $x_i$ and $y_j$. We will see in the
following that this property implies both squeezing and the
product rule for FQH wave functions.

\subsection{Spinless FQH wave functions and parafermionic CFTs}

\subsubsection{Parafermionic CFTs}

Many fully polarized QH wave functions can be written in terms of
parafermionic CFTs. This includes Laughlin~\cite{laughlin83prl1395},
Moore-Read~\cite{Moore-91npb362} and the Read-Rezayi states
\cite{Read-99prb8084} (as well as all Jack states
\cite{bernevig-08prl246802}), but also any generalized parafermionic
states, such as $N=1$ superconformal or $S_3$ wave functions
\cite{estienne-10npb539,simon-10prb121301}.

Here we constrain ourselves to a condensed derivation required to
obtain the product rule from CFT. More details about parafermions in
the context of FQHE can be found in
Ref.~\onlinecite{estienne-10npb539}.  These parafermionic CFTs,
denoted as $\mathbb{Z}_k^{(r)}$, contain a set of $k$ parafermionic
primary fields $\{ \Psi_0 = 1, \Psi_1, \cdots \Psi_{k-1} \}$ with the
following fusion rules:
\begin{equation}
\Psi_n \times \Psi_m \to \Psi_{n+m} \quad \textrm{mod } k
\end{equation}
The conformal dimension of the field $\Psi_n$ is $\Delta_n =
\frac{r}{2}\frac{n(k-n)}{k}$, and its conjugate is $\Psi_{k-n}$.

Parafermionic FQHE wave functions take the following form:
\begin{equation}
P(z_1, \cdots, z_N) = \langle \Psi(z_1) \cdots \Psi(z_N) \rangle \prod_{i<j} (z_i - z_j)^{r/k}.
\end{equation}

Using \eqref{generic_factorization} for $N=N_A+N_B$  in the case in which all the primary fields $\Phi_i$ are taken to be
parafermionic fields $\Psi$ in a $\mathbb{Z}_k^{(r)}$ theory, we have:
 \begin{eqnarray}
\lefteqn{\gamma^{N_B\Delta_1 + \Delta_a}C(x_i | \gamma y_j)   \sim } \nonumber \\ 
& & \langle \Psi (x_1) \cdots \Psi(x_{N_A}) \Psi_{-a}(\infty) \rangle
\langle \Psi_{a}(0) \Psi(y_1)\cdots \Psi( y_{N_B}) \rangle \nonumber \\ 
\end{eqnarray}
where $a = N_A\; \text{mod} \;k$ corresponds to the sector in which the $N_A$
particles live after the cut, and $\Delta_a =
\frac{r}{2}\frac{a(k-a)}{k}$.  Equivalently, for the wave function this reads
$P(x_i|y_j) = C(x_i | y_j) \prod (x_i-x_j)^{\frac{r}{k}} \prod
(y_i-y_j)^{\frac{r}{k}} \prod (x_i-y_j)^{\frac{r}{k}}$:
 \begin{eqnarray}
\lefteqn{\gamma^{-\frac{r}{2k}( 2N_AN_B + N_B (N_B-k) - a(k-a))} P_{N}(x_i|\gamma y_j) \sim } \nonumber \\
& &    P^{(a)}_{N_A}(x_i) P^{(\bar a)}_{N_B}(-1/y_j)  \left(\prod_{i} y_i
\right)^{\frac{r}{k}(N-k)}. \label{factorization}
\end{eqnarray}
In particular, in the neutral sector $a=0$ we have
\begin{equation}
P_{N}(x_i|\gamma y_j) \sim \gamma^{N_B N_{\Phi}(N_B)/2} \prod_{i} (\gamma y_i)^{r N_A/k} P_{N_A}(x) P_{N_B}(y),
\end{equation}
where we introduced $N_{\phi}(N) = r(N-k)/k$.

\subsubsection{Squeezing and product rule}
From~\eqref{factorization} one can derive the following properties:

{\it Squeezing.}
Consider the power of $\gamma$ in \eqref{factorization} for $N_B=1$: 
\begin{equation}
P_{N}(z_1,z_2,\cdots z_{N-1},\gamma z_N) \sim_{\gamma \to \infty} \gamma^{r(N-k)/k} . 
\end{equation}
This shows that for any monomial $m_{\mu}$ entering the decomposition of
$P_N$ we have $\mu_1 \leq \lambda_1 = \frac{r}{k}(N-k)$.  By iteration
on $N_B$ in~\eqref{factorization}, one finds that (i) the root
partition is $\lambda=(k 0^{r-1} k 0^{r-1} \cdots k)$ and (ii) any other partition $\mu$ obeys $\mu_1+ \cdots + \mu_i
  \leq \lambda_1 + \cdots + \lambda_i$, \emph{i.e.} is obtained by
  squeezing from $\lambda$.

{\it Product rule.}
In the limit we consider we send the $N_B$ particles to infinity:
\begin{equation}
\lim_{\gamma \to \infty} \gamma^{-\frac{r}{2k}( 2N_AN_B + N_B (N_B-k) - a(k-a))}P_{N}(x_i | \gamma y_j)
\end{equation}
only the monomials $m_{\mu}$ such that 
\begin{equation}
\mu_1 + \cdots \mu_{N_{B}} = \lambda_1 + \cdots \lambda_{N_{B}} 
\end{equation}
survives. This kills all monomials obtained by squeezing through the cut
between $N_A$ and $N_B$ particles, and leaves the others invariant.  Now
that squeezing has been established, the product rule is simply
equivalent to the monomial decomposition of the factorization
property:
 \begin{eqnarray}
\lefteqn{\lim_{\gamma \to \infty} \gamma^{-(\lambda_1 + \cdots + \lambda_{N_B})}P_{N}(x_i | \gamma y_j) = } \nonumber \\
& &  P^{(a)}_{N_A}(x_i) \tilde{P}^{(\bar a)}_{N_B}(y_j)  \left(\prod_{j=1}^{N_B} y_j \right)^{rN_A/k}  
\end{eqnarray}
where $\tilde{P}_{N_B}$ stands for the north (south) pole reflection
of $P_{N_B}$.

\subsection{Spin singlet states}

The same argument from above applies to spin-unpolarized states, such as the
Haldane-Rezayi~\cite{haldane-88prl956}, Halperin
\cite{halperin83hpa75}, and NASS states~\cite{ardonne-99prl5096}. Their
CFT description involves several types of electron operators,
typically consisting of a parafermion field and a vertex operator of a
set of chiral boson fields. These boson fields $\Phi_c$ and $\Phi_s$ usually describe
charge and spin associated with the particles. The electron creation
operators take the generic form:
\begin{eqnarray}
V_{\uparrow}(z) &= &\Psi_{\uparrow}(z) :e^{\frac{i}{\sqrt{2}} \left( \sqrt{\beta+\gamma} \Phi_c + \sqrt{\beta-\gamma}\Phi_s\right)}: , \\
V_{\downarrow}(w) &= &\Psi_{\downarrow} (w):e^{\frac{i}{\sqrt{2}} \left( \sqrt{\beta+\gamma} \Phi_c- \sqrt{\beta-\gamma}\Phi_s\right)}:,
\end{eqnarray}
where $\Psi_{\uparrow}$ and $\Psi_{\downarrow}$ can be trivial fields
(Halperin), Gepner parafermions (NASS)~\cite{gepner} or ghosts
(Haldane-Rezayi)~\cite{gurarie-97npb513}. The values of the rational numbers $\beta$ and $\gamma$ entering the vertex operators depend on this CFT.

The spin polarized wave
function assumes the form
 \begin{eqnarray}
\lefteqn{
P(z_i,w_j) = \langle  \Psi_{\uparrow}(z_1)\Psi_{\downarrow}(w_1) \cdots \Psi_{\uparrow}(z_N)\Psi_{\downarrow}(w_N) \rangle } \nonumber \\
& & \prod_{i<j} (z_i-z_j)^{\beta}  (w_i-w_j)^{\beta} \prod_{i,j}(z_i-w_j)^{\gamma}.
\end{eqnarray}
Taking the asymptotic factorized behavior as $n$ up spin and $m$
down spin electrons are taken to infinity as in
\eqref{generic_factorization}, we first obtain a {\it weak form of squeezing}: there exists an integer $N(n,m)$
  dominating all the partitions $(\mu^{\uparrow},\mu^{\downarrow})$ in
  the sense that
\begin{equation}
\mu_1^{\uparrow} + \cdots \mu_n^{\uparrow} + \mu_1^{\downarrow} + \cdots \mu_m^{\downarrow}  \leq N(n,m),
\end{equation}
which is due to the asymptotic behavior $P \sim \gamma^{N(n,m)}$. This
integer $N(n,m)$ can be expressed in terms of the CFT data $\alpha,
\beta, \Delta_{\uparrow}, \Delta_{\downarrow}$ and $\Delta_a$.

Second, we obtain a {\it product rule} if a partition $\mu = (\mu_{\uparrow},
  \mu_{\downarrow}) $ is separable, \emph{i.e.} $\mu = \mu_A + \mu_B$.
  With $\mu_A$ ($\mu_B$) squeezed from $\lambda_A$ ($\lambda_B$),
  the corresponding monomial (slater) coefficient is $m_{\lambda \mu}
  = m_{\lambda_1 \mu_1} \times m_{\lambda_2 \mu_2} $.
In order to obtain a stronger form of squeezing, a more detailed
analysis is involved.  One would have to specify the OPEs of the
operators $\Psi_{\uparrow}$ and $\Psi_{\downarrow}$, and in particular
work out the dimension of the field $\Phi_a$ appearing in the fusion
of $n$ fields $\Psi_{\uparrow}$ and $m$ fields $\Psi_{\downarrow}$, as
we did for the spin-polarized case. The detailed analysis of the NASS
and Halperin states is beyond the scope of this paper. We treat the HR state in Appendix \ref{HR_CFT}.

\section{Conclusion}
\label{sec:con}

In this paper we have given an extended derivation of a recurrence
formula for the Slater decomposition of fermionic Jack polynomial
states.  We have given a rigorous and detailed account on the product
rule symmetry for spin-polarized quantum Hall trial states first
presented in Ref.~\onlinecite{bernevig-09prl206801}. We generalized
the whole approach to spinful states and specifically derived a
recurrence relation for the spinful slater determinant decomposition
of the Haldane Rezayi state for which we have computed its spinful
geometric entanglement spectrum. The product rule symmetry is found to
be a deep general property of quantum Hall trial states, involving
both fermionic and bosonic as well as spinful and spin-polarized
states. While for states described by parafermionic conformal field theory
(which include but transcend the Jack polynomials) we were able to
show that the product rule comes out of the parafermionic fusion
properties, for many other FQH states the product rule holds even
though the states cannot be described by analogue conformal field theory.
We showed that the product rule can be used as an increasingly good
approximation of the FQH state. 

While the product rule symmetry is not exactly valid for ground states
built from realistic interactions (such as the Coulomb potential), we
would like to investigate if an approximated version of this symmetry
also manifests itself in these cases. Such property would greatly help
to improve any density matrix renormalization group algorithm designed
for fractional quantum Hall systems.



\begin{acknowledgments}
  We thank F.D.M. Haldane, R. Santachiara, K. Schoutens, and S. Simon
  for valuable discussions. RT, BE, and BAB thank the Ecole normale
  superieure (Paris) for hospitality. RT and BAB thank the center for
  international collaborations in Beijing where parts of this work
  have been done. RT is supported by a Feodor Lynen Fellowship of the
  Humboldt Foundation and Alfred P. Sloan Foundation funds. BE was
  supported by FOM of the Netherlands. NR was supported by Agence
  Nationale de la Recherche under Grant No. ANR-JCJC-0003-01.  BAB
  acknowledges Princeton University startup funds, Alfred P. Sloan
  Foundation funds, and NSF CAREER DMR-0952428.
\end{acknowledgments}  

\begin{appendix}
\section{Helpful Formulas}
\label{app:form}
We present several formulas used to simplify certain polynomial terms.
They follow from elementary algebra and are easily derived by
exploiting the symmetry with respect to the summation indices.  In
Sec.~\ref{sec:fermion}, we use
\begin{equation}
\sum_{i \ne m}^N \frac{z_i}{z_i-z_m}= \frac{1}{2} N(N-1),
\end{equation}

\begin{equation}
\sum_{\substack{i,m,n \\ i \ne j \ne m \ne i}} \frac{z_i^2}{(z_i-z_m)(z_i-z_n)} = \frac{1}{3} N(N-1)(N-2).
\end{equation}

In Sec.~\ref{sec:hr} and Appendix~\ref{app:hr}, we use

\begin{eqnarray}
&&\sum_{i\; k\ne l} \frac{z_i^2}{(z_i-w_k)(z_i-w_l)}+\sum_{j\ne i;k}
\frac{z_i z_j-w_k(z_i-z_j)}{(z_i-w_k)(z_j-w_k)} \nonumber \\
&&+ \sum_{i; k\ne l}\frac{w_i^2}{(w_i-z_k) (w_i-z_l)} +\sum_{j\ne i ;
  k} \frac{w_i w_j - z_k (w_i+w_j)}{(w_i-z_k)(w_j-z_k)}\nonumber \\
&&=2N^2(N-1).
\label{form-hr}
\end{eqnarray}

\section{Example for Monomial decomposition}
\label{app:mono}
\begin{figure*}[h]
  \begin{minipage}[l]{0.99\linewidth}
    \includegraphics[width=\linewidth]{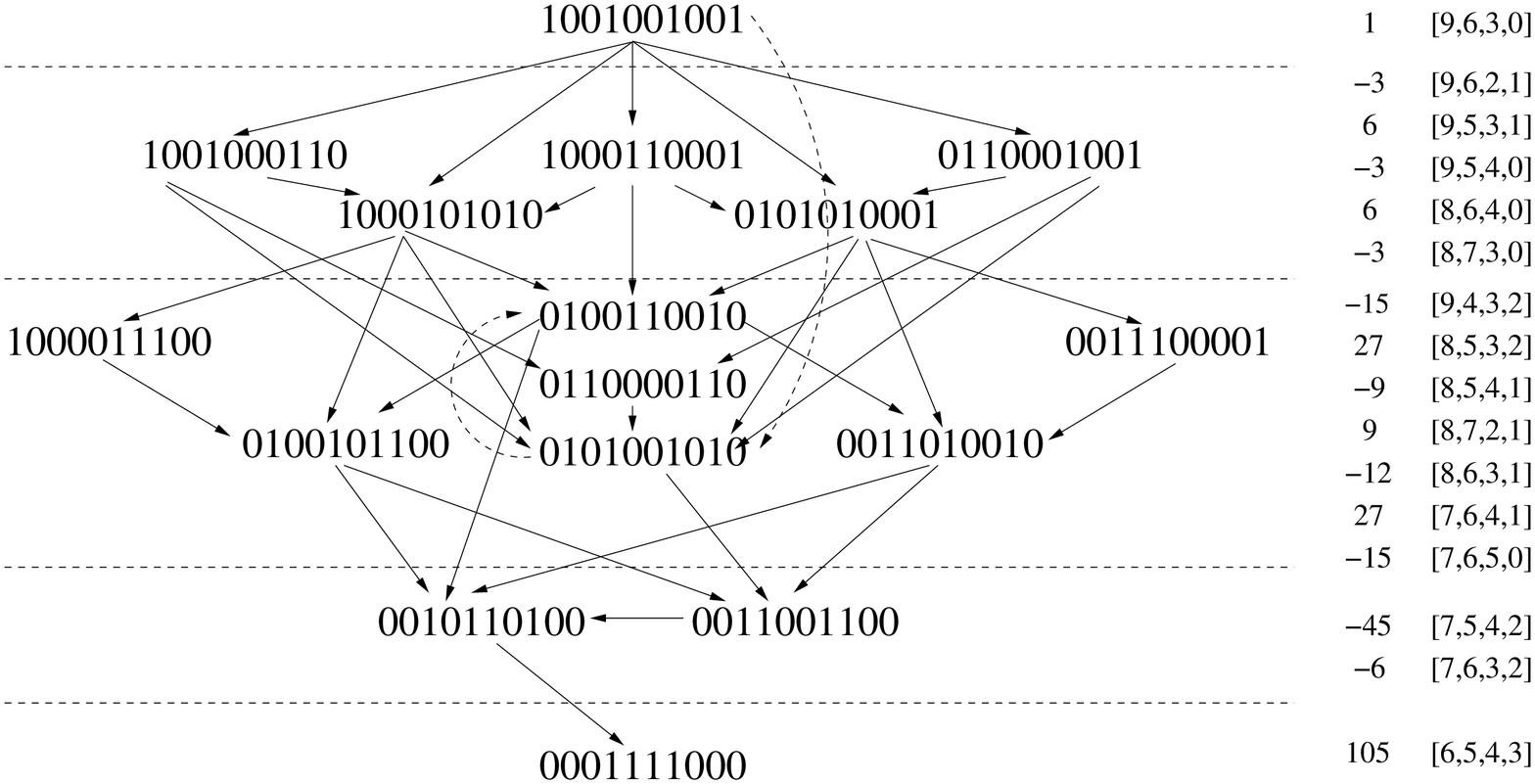}
  \end{minipage}
  \caption{Monomial decomposition of $N=4$ particle $\nu=1/3$ Laughlin
    state.  The particle positions are denoted by $1$. The
    coefficients of the Slater partitions are computed according
    to~\eqref{slater1}, with $\alpha=-2$. The arrows denote a
    squeezing relation from the upper to the lower partition. In
    total, there are 4 squeezing levels till the maximally squeezed
    partition is reached. }
\label{monolaughlin}
\vspace{-0pt}
\end{figure*}
In Fig.~\ref{monolaughlin} we give an example of how to use the
squeezing and product rule properties to obtain the $N=4$-particle
$\nu=1/3$ Laughlin state.
\section{Fermionic Laplace-Beltrami operator}
\label{app:flb}
We want to rephrase the derivatives on $J_{\lambda_B}^{\alpha}$ as a
derivative action on $S_{\lambda}^\alpha$. This can be done in a
compact form since they only differ by a Jastrow factor
multiplication. The first derivative yields:
\begin{equation}
z_i \frac{\partial}{\partial z_i}S_{\lambda}^{\alpha} = \left[ z_i \frac{\partial}{\partial z_i} J_{\lambda_B}^{\alpha} \right] \prod_{k<l} (z_k-z_l) + \sum_{\substack{m \\ i\ne m}} \frac{z_i}{z_i-z_m} S_{\lambda}^{\alpha}.
\end{equation}

Similarly, plugging in the previous result for the first derivative,
the second derivative action can be rewritten as:

\begin{eqnarray}
&&\left(z_i \frac{\partial}{\partial z_i} \right)^2 S_{\lambda}^\alpha = \left[ \left( z_i \frac{\partial}{\partial z_i} \right)^2 J_{\lambda_B}^{\alpha}\right] \prod_{k<l} (z_k-z_l) \nonumber \\
&&+\sum_{\substack{m \\ i\ne m}}\frac{2z_i^2}{z_i-z_m} \frac{\partial}{\partial z_i} S_{\lambda}^\alpha + \left[ \sum_{\substack{m\\ i\ne m}}  \frac{z_i}{z_i-z_m} -\frac{z_i^2}{(z_i-z_m)^2} \right. \nonumber \\
&& \left. - \sum_{\substack{m,n \\ m\ne i, n\ne i}} \frac{z_i^2}{(z_i-z_m)(z_i-z_n)}\right] S_{\lambda}^\alpha.
\end{eqnarray}

\section{Haldane Rezayi state}
\label{app:hr}
\subsection{Derivative action}
Starting from the CFT differential equation for the non-unitary
term~\eqref{eq:hr-nu1}, we perform a calculation similar to the
previous derivation of the fermionic LB operator.
For generality, we can consider the generalized Haldane-Rezayi state:
\begin{equation}
\Phi_{\text{HR}}^{m,m,n} = \Phi_0 \prod_{i<j}(z_i-z_j)^m \prod_{i<j}(w_i-w_j)^m \prod_{i,j}(z_i-w_j)^n.
\end{equation}
Following a tedious, but straightforward calculation, we find
\begin{widetext}
\begin{eqnarray}
&&\left[ \frac{1}{2} \sum_i \left(z_i \frac{\partial}{\partial z_i} \right)^2 -\frac{1}{2} \sum_i z_i \frac{\partial}{\partial z_i} -\sum_{i\ne j}(z_i+z_j) \frac{\partial}{\partial z_j} -(m-1) \sum_{j \ne i} \frac{z_i^2}{z_i-z_j}\frac{\partial}{\partial z_i}- n \sum_{i,k} \frac{z_i^2}{z_i-w_k}\frac{\partial}{\partial z_i}-\sum_{i,k} \frac{z_i^2}{z_i-w_k}\frac{\partial}{\partial w_k}\right.\nonumber \\
&&+\frac{m^2-m-2}{2} \sum_{i\ne j} \frac{z_i^2}{(z_i-z_j)^2}+\frac{n^2-n-2}{2}\sum_{k,i} \frac{z_i^2}{(z_i-w_k)^2} +\frac{n^2+m}{2}\sum_{i; k\ne l} \frac{z_i^2}{(z_i-w_k)(z_i-w_l)} \nonumber \\
&&\left. + \frac{n(m+1)}{2}\sum_{j\ne i ; k} \frac{z_i z_j -w_k(z_i+z_j)}{(z_i-w_k)(z_j-w_k)}+ \frac{m(m+1)}{6}N(N-1)(N-2)\right]\Phi_{\text{HR}}^{m,m,n}=0.
\end{eqnarray}
\end{widetext}
One observes that the $m=2, n=2$ case is special: the three-body terms
have an identical coefficient equal to $3$. These 3-body terms reduce
to a constant once the above equation for $z$ is added to the similar
equation for $w$. By use of~\eqref{form-hr}, this leads to the
expression~\eqref{eq-hr}.
\subsection{Recurrence formula}
\subsubsection{Equal spin action}
From the differential equation~\eqref{eq-hr}, we know that under the
action of the equal and different-spin terms, the sum
over Slaters contained in the HR state must yield another sum of
Slaters.  Thus, to deduce a decomposition formula we enforce this condition. Without loss of generality, we
can consider the operator action on a two-particle Slater state, and
explicitly demand
\begin{eqnarray}
&&\frac{z_1+z_2}{z_1-z_2}\left(z_1 \frac{\partial}{\partial z_1} - z_2\frac{\partial}{\partial z_2} \right) \sum_{n_1,n_2}a_{n_1,n_2}\text{sl}_{n_1,n_2}\nonumber\\
&&=\sum_{n_1,n_2}b_{n_1,n_2}\text{sl}_{n_1, n_2},
\label{eespin}
\end{eqnarray}
where we use $n_1$ ($n_2$) as momentum indices of the 
two-particle same-spin Slater determinants.  Working out the derivative action
and expanding the expression inside the equality, we find
\begin{eqnarray}
&&\sum_{n_1,n_2}\left(a_{n_1,n_2} (n_1-n_2)+a_{n_1+1,n_2-1}(n_1-n_2+2) \right)\nonumber \\
&&\left(z_1^{n_1+1}z_2^{n_2}+z_2^{n_1+1}z_1^{n_2}\right) =\nonumber \\
&&\sum_{m_1,m_2} \hspace{-4pt} (b_{m_1,m_2}-b_{m_1+1,m_2-1})( z_1^{m_1+1}z_2^{m_2}+z_2^{m_1+1} z_1^{m_2}).\nonumber \\
\end{eqnarray}
This equation can be worked out iteratively, starting from the maximum
polynomial degree. In terms of the $b$ coefficients in the final
Slater superposition, the coefficients $a$ are given by
\begin{equation}
b_{\mu_1,\mu_2}=(\mu_1-\mu_2)a_{\mu_1,\mu_2} + 2 \sum_i a_{\mu_1+i, \mu_2-i} (\mu_1-\mu_2+2i).
\end{equation}
Since the interaction is pairwise, this can be directly generalized to
larger particle numbers of Slaters, and finally yields
expression~\eqref{equal-big}.

We now illustrate that the degree of the Jastrow factor in the
polynomial determines whether the action of Eq.~\ref{eespin} causes
fractions or not. As an example, we consider the
$\text{sl}_{3,0}=z_1^3-z_2^3$ two-particle Slater determinant. The
action of the operator in Eq.~\ref{eespin} gives:
\begin{equation}
\frac{z_1+z_2}{z_1-z_2}\left(z_1 \frac{\partial}{\partial z_1} - z_2\frac{\partial}{\partial z_2} \right) \text{sl}_{3,0}= 3 \frac{z_1+z_2}{z_1-z_2} (z_1^3+z_2^3),
\end{equation}
which cannot be decomposed into polynomials without fractions. As
stated before, this case does not occur for the HR state, since there
is an equal-spin Jastrow factor of $2$nd power. Let us now consider
the polynomial which is constructed from $\text{sl}_{3,0}$ times
a Jastrow factor:
\begin{eqnarray}
&&\frac{z_1+z_2}{z_1-z_2}\left(z_1 \frac{\partial}{\partial z_1} - z_2\frac{\partial}{\partial z_2} \right) \text{sl}_{3,0}(z_1-z_2)\nonumber \\
&=& \frac{z_1+z_2}{z_1-z_2}(3(z_1^3+z_2^3)(z_1-z_2)+z_1 z_2 (z_1^2-z_2^2)\nonumber\\
&& + z_1^4-z_2^4).
\end{eqnarray}
Since all terms from above cancel the $z_1-z_2$ fraction, the
polynomial composed out of a Slater times a Jastrow factor gives no
fractions upon the action of the operator in
Eq.~\ref{eespin}. This holds for any higher power of Jastrow factors
multiplied with the Slater determinant.

\subsubsection{Inter-spin action}
To compute the action on the inter-spin term, we again constrain
ourselves to a two-particle mixed-spin Slater partition $z_1^n w_1^m +
z_1^m w_1^n$, and assume $n \ge m$ (we remember that the $m \ge
n$ Slater has the same coefficient):

\begin{eqnarray}
&&\frac{z_1+w_1}{z_1-w_1}\left( z_1\frac{\partial}{\partial z_1} -w_1\frac{\partial}{\partial w_1} \right) (z_1^n w_1^m+z_1^m w_1^n)\nonumber \\
&&=(n-m)(z_1^n w_1^m+z_1^m w_1^n) \nonumber \\
&&+ 2(n-m) \sum_{i=1}^{(m-n)/2-1} (z_1^{n-i}w_1^{m+i}+z_1^{m+i}w_1^{n-i}) \nonumber \\
&&+2(n-m)z_1^{(n+m)/2}w_1^{(n+m)/2}.
\label{appnonequal}
\end{eqnarray}
For $(n-m)\equiv 1\text{mod}2$, the upper limit of the sum is a half
integer. In this case, the sum is evaluated as an analytical extension
from an integer to a half integer upper limit. To each of these
Slaters, we also add the counterpart $n\leftrightarrow m$. To prevent
double counting, whenever $m=n$ the counterpart must not be added. In
the language of single Slaters, this demands the division of the
equal-momentum Slater prefactor by 2 in comparison to other Slaters.
For the general particle case, this yields the following recipe: Sum
over all combinations of pairs $(i,k)$ with one coordinate from the
$\uparrow$ spin partition and one from the $\downarrow$ spin
partition. If $\lambda_i > \lambda_k$ ($n>m$ as above) take the
prefactor as given in~\eqref{appnonequal}. If $\lambda_i < \lambda_k$,
change $\uparrow \leftrightarrow \downarrow$ and add it (the
spin-rotated Slater must have the same coefficient). Produce all
allowed squeezings and add the Slater terms.  If one pair $(i,k)$ and
some squeezed partition produces an $n=m$ state (i.e. a doubly
occupied single particle state), and the remainder $\uparrow$ spin and
$\downarrow$ spin partition are equivalent as well, this factor must
be divided by two (or rather, in our case, changed from $2$ to $1$).
We then obtain~\eqref{eq-unequal}.
\section{Example for spinful Slater decomposition}
\label{app:sldec}
\begin{figure*}[h]
  \begin{minipage}[l]{0.99\linewidth}
    \psfrag{a}{$0$}
    \psfrag{b}{$\uparrow$}
    \psfrag{c}{$\downarrow$}
    \includegraphics[width=\linewidth]{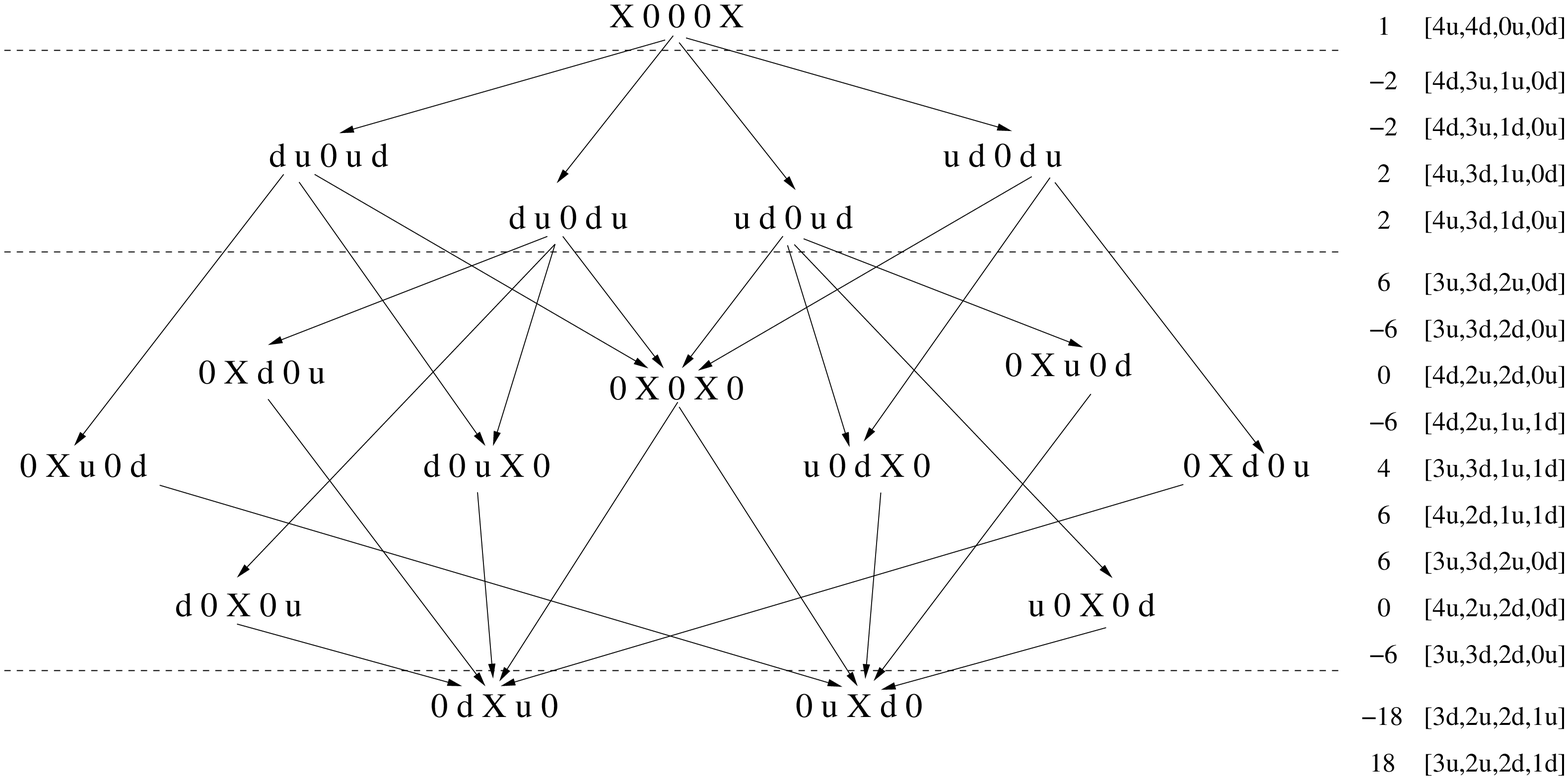}
  \end{minipage}
  \caption{Spinful Slater decomposition of the $N=4$ Haldane-Rezayi
    state.  The notation corresponds to $X=\uparrow \downarrow$ on the
    same orbital, and $u\equiv \uparrow$, $d\equiv \downarrow$. The
    spinful partitions are written in spin-mixed fashion and ordered
    with respect to momentum. The coefficients are computed
    corresponding to Eq.~\ref{rechr}. The two appearing zero weight
    partitions are a manifestation of the zero-weight rule according
    to~\eqref{p1}. The arrows denote that a given partition can be
    squeezed from the partition above. There are three squeezing
    levels until the maximally squeezed partitions are reached.}
\label{monohr}
\vspace{-0pt}
\end{figure*}
In Fig.~\ref{monohr}, we give an example of the HR state decomposed
into squeezed partition from the root partition. We show the full
decomposition of the $N=4$ HR state into spinful partitions squeezed
from $X000X$. We observe an example of a partition squeezed from
$X000X$ that has zero coefficient, i.e.  $\text{u}0X0\text{d}$: it cannot be
constructed from $P_1=X0X$ times $101$ for $\uparrow$ spins and $101$
for $\downarrow$ spins.

\section{Haldane-Rezayi, CFT analysis}
\label{HR_CFT}

As mentioned before in Section~\ref{sec:hr}, the CFT corresponding to
the HR state is a logarithmic theory that contains two fields $1$ and
$\tilde{1}$ with vanishing conformal
dimension~\cite{gurarie-97npb513}. This particularity makes the CFT
treatment more complicated. Fortunately, the factorization can already
be obtained from the explicit form of the correlation function:
 \begin{eqnarray}
\lefteqn{
\langle \partial\theta(z_1)  \partial\bar{\theta}(w_1) \cdots \partial \theta(z_N)  \partial\bar{\theta}(w_N) \tilde{I} \rangle = } \nonumber \\
& &  \sum_{\sigma} \epsilon(\sigma) \prod_i \frac{1}{(z_i-w_{\sigma(i)})^2}.
\end{eqnarray}
The r.h.s. trivially obeys the following asymptotic behavior as $N_B$
up-spins and $N_B$ down-spins are taken to infinity:
 \begin{eqnarray}
\lefteqn{
\langle  \partial \theta(z_1) \partial \bar{\theta}(w_1) \cdots   \partial\bar{\theta}(w_{N_A}) \partial \theta(\gamma z_{N_A+1})  \cdots   \partial\bar{\theta}(\gamma w_{N_A+N_B}) \rangle \sim } \nonumber \\ 
& & \gamma^{-2N_B}  \langle \partial \theta(z_1) \cdots \partial
\bar{\theta}(w_{N_A}) \rangle \langle \partial \theta(z_{N_A+1})
\cdots \partial \bar{\theta}(w_{N_A+N_B}) \rangle. \nonumber \\
\end{eqnarray}
This yields the product rule (for a cut through an empty orbital)
using the same steps as in the spin-polarized case. Cuts through an
occupied orbital can be analysed in a similar way, by looking at the
asymptotic behavior as $N_B$ up-spins and $N_B+1$ down-spins are sent to
infinity.

\end{appendix}


\end{document}